
        \input amstex
        \parskip=8pt plus 1pt minus 1pt
        \baselineskip=18pt
        \magnification\magstephalf
        \TagsOnRight
        \def \a{\alpha}
        \def \G{\Gamma}
        \def \g{\gamma}
        
        \def \de{\delta}
        \def \De{\Delta}
        \def \ep{\varepsilon}
        \def \la{\lambda}
        \def \La{\Delta N}
        \def \r{\rho}

        \def \om{\omega}
        \def \mat{\mu_\a(dx;z)}
        \def \matt{\mu_\a(dt;z)}
        \def \ptz{p_\a(t;z)}
        \def \Var{\text{\rm Var}\;}
        
        \def \f{\varphi}
        \def \N{\bold N}
        \def \Q{\bold Q}
        \def \R{\bold R}
        \def \C{\bold C}
        \def \T{\bold T}
        \def \Z{\bold Z}
        
        \def \mes{\,\text{\rm mes}\,}
        \def \Id {\,\text{\rm Id}\,}
        \def \lT{\limsup_{T\to\infty}}
        \def \liT{\lim_{T\to\infty}}
        \def \lLT{\lim_{S/T\to\infty,\,S/T^2\to 0}}
        \def \lLt{\lim_{T\to\infty,\,S/T\to z}}
        \def \lL0{\lim_{S\to\infty,\,S/T\to 0}}
        \def \lt0{\lim_{T\to\infty,\,S/T^2\to 0}}
        \def \lN{\lim_{N\to\infty}}
        
        \def \lc{\lim_{\Delta c\to 0}}
        \def \ls{\limsup}
        \def \di{\displaystyle}
        \def \IT0{{1\over T}\int_{T_0}^T}
        \def \IT{{1\over T}\int_1^T}
        \def \IO{{1\over T}\int_0^T}
        
        \def \IC{{1\over(c_2-c_1)T}\int_{c_1T}^{c_2T}}
        \def \IDC{{1\over T\Delta c}\int_{T}^{T(1+\Delta c)}}
        \def \Ic{{1\over(c_2-c_1)}\int_{c_1}^{c_2}}
        \def \IB{\IC}

        \def \SZ{\sum_{n\in\bold Z^2\setminus\{0\}}}

        \def \tf{{3\pi\over 4}}

        \def \NS{|n|^{-3/2}\sqrt{\rho(n)}}
        \def \NSS{|n|^{-3}\rho(n)}

        \def \CS{\cos(2\pi RY(n)+\phi(n;\a))}

        \def \PI{\pi^{-1}}

        \def \ERL{\La (R,S;\a)}
        \def \FR{F(R;\alpha)}
        \def \ER{\La (R;\alpha)}
        \def \FRL{F(R,S;\a)}
        \def \FW{F(R+\wL;\a)}

        \def \II{\int_{-\infty}^\infty}

        \def \Rw{R+\wL}
        \def \sz{\sum_{n\in\Z^2\setminus\{0\}}}
        
        \def \Sl{\sum_{l=1}^\infty}
        
        \def \Sk{\sum_{k=1}^\infty}
        
        \def \wL{w_S(R)}

        \def \LC{L^\infty([c_1,c_2])}
        \def \dR{\,\f(R/T)\,dR}

\def \fs{f_n(s;z,\a)}
\def \fk{f_k(s;z,\a)}
\def \Cb{C_{\text{\rm b}}(\R^1)}

        \null

\rightline{IASSNS-HEP-93/14}

\rightline {March 1993}

\vskip 3mm

        \centerline {\bf ENERGY--LEVEL STATISTICS OF MODEL QUANTUM SYSTEMS:}
        \vskip 2mm
        \centerline   {\bf UNIVERSALITY AND SCALING IN A LATTICE--POINT
PROBLEM}
        \vskip 20mm

        \centerline{\bf Pavel M. Bleher}

        \vskip 1mm

        \centerline {School of Natural Science }

        \vskip 1mm

        \centerline {Institute for Advanced Study }

        \vskip 1mm

        \centerline{Princeton, NJ 08540, USA}

        \vskip 1mm

        and

        \vskip 1mm

        \centerline{\bf Joel L. Lebowitz}

        \vskip 1mm

        \centerline{Departments of Mathematics and Physics}

        \vskip 1mm

        \centerline{Rutgers University}

        \vskip 1mm

        \centerline{New Brunswick, NJ 08903, USA}

\vskip 1cm

{\bf Abstract.} We  investigate the statistics of the number
$N(R,S)$ of lattice points, $n\in \Z^2$, in a ``random''
annular domain $\Pi(R,w)=\,(R+w)A\,\setminus RA$,
where $R,w >0$.
Here $A$ is a fixed convex set with smooth boundary and
$w$ is chosen so that the area of $\Pi (R,w)$ is $S$.
The randomness comes from $R$ being taken as random ( with a smooth
denisity ) in some interval $[c_1T,c_2T]$, $c_2>c_1>0$.
We find that in the limit $T\to\infty $ the variance and
distribution of $\De N=N(R;S)-S$ depends strongly on how
$S$ grows with $T$. There is a saturation regime $S/T\to\infty$,
as $T\to\infty$ in which the fluctuations in $\Delta N$ coming
from the two boundaries of $\Pi $, are independent.
Then there is a scaling regime, $S/T\to z$, $0<z<\infty $
in which the distribution depends on $z$ in an almost
periodic way going to a Gaussian as $z\to\ 0$. The variance
in this limit approaches $z$ for ``generic'' $A$ but can be
larger for ``degenerate'' cases. The former behavior is
what one would expect from the  Poisson limit of a
distribution for annuli of finite area.

        \vfill\eject
        \beginsection 1. Introduction and Formulation of Results \par

         This paper continues our study of the distribution of
integer lattice points inside
        a ``random'' region on the plane
(see [BCDL], [CL], [Bl2], [Bl3], [BD1] and [BD2]).
While the question can be thought of as primarily number theoretical,
our motivation  comes from
the interest in the distribution of eigenvalues of quantum systems.
The later problem has received a great
deal of attention in recent years  with particular emphasis on the following
que
   stion:
how does the statistics of the eigenvalues relate to the nature
of the corresponding classical dynamical system?

        One of the striking conjectures
        is a universality of the local statistics of eigenvalues
        of generic quantum Hamiltonians:
 for integrable systems the local statistics
is  Poissonian, while for
        hyperbolic systems it is the Wigner statistics
        of the ensemble of Gaussian matrices.
This conjecture is based on a number of analytical,
        numerical and experimental results (see, e.g., [B1], [Boh], [Gu], [GH])
         and the task of  theory is to explain them and
        to find out their range of validity.
To see the connection between the lattice--point problem and the statistics
        of eigenvalues in the integrable case, consider
        a  simple model system, a free particle in a
        rectangular box with periodic boundary conditions. In this case
        the eigenvalues are
        $$
        E_n={n_1^2\over {a_1^2}}+ {n_2^2\over {a_2^2}}, \quad
        n=(n_1,n_2)\in \Z^2,
        $$
        where $2\pi a_1,2\pi a_2 $ are the sides of the box.
        The problem then is:
        what are the statisics of the numbers
         $E_n$?  This is clearly the same problem as finding the statistics of
l
attice
points inside a suitable ellipse. More generally we may consider integrable
syst
   ems with eigenvalues
$$
E_n\,=\,I(n_1-\a_1,n_2-\a_2),\quad n=(n_1,n_2),\eqno (1.1)
$$
where $I(x_1,x_2)$ is a homogeneous function of second order and ask about the
s
   tatistics of $E_n$.
Now, in the sequence $\{E_n,\, n\in \Z^2\, \} $
        there is nothing random, so the first question is, what
do we mean by statistics of the sequence $E_n$?
        To describe it let us consider the sequence of energy levels $E_n$
        in the interval  $[E,E+S]$ with $E\gg S\gg 1$
(the average spacing between levels is of order of $1$).
        Assuming that $E$ is uniformly distributed on an interval
$c_1T \le E \le c_2 T $,
        we may consider the
        sequence $ X_E =\,\{\, E_n-E \, : \, E \le E_n < E+S \} $
        as a random one.
        The question is: does the limit distribution of $X_E$ exist,
        when $T,S\to\infty $ and $S$ is a prescribed function of $T $. Is this
        limit distribution Poisson?

        This problem was considered originally in the work
        of Berry and Tabor [BT] (see also the review paper [B1]),
where convincing
        physical arguments were presented in  favor of a Poisson
        limit distribution.
In particular the distribution of the distances
        between neighboring levels was found numerically to be
        exponential which fits to  Poisson statistics.
        Sinai in [S1] and Major in [M] (see also [Bl1]) studied rigorously
        the Poisson limit distribution in a model lattice
        problem. They showed that for a typical (in a probability
        sense) oval in a plane the number of integral points
        in  random narrow strip of a fixed area
        between two enlarged ovals has a Poisson distribution.
        Major proved also some other results in this direction,
        and he showed that a typical oval from the probability
        spaces, which were used in his work and in the work
        of Sinai, does not belong to $C^2$.
        For typical smooth, say $C^4$--, ovals the Poisson conjecture,
        for the number of integral points in a random narrow
        strip of a fixed area, remains open; see [CLM] for some related
results.

        In a different direction  Casati, Guarneri and Valz-Gris [CGV]  argued
t
hat
        the sequence of  levels cannot be
        considered as truly random.
In fact, Casati, Chirikov and Guarneri  (see [CCG]) found numerically
        a saturation of rigidity at large energies, which
        gave an estimate of the range of the applicability of
        the Poisson conjecture (the rigidity is a statistical characteristic
        which estimates how well the counting function of the
        energy levels is approximated locally by a linear
        function; it was introduced by Dyson and Mehta
        in their studies of  ensembles of Gaussian
        matrices, see [Me]).

        Berry carried out an analytical analysis of the
        saturation of the rigidity in  [B2].
        He showed  that
        the rigidity has a crossover at the scale of $S$
        of the order $E^{1/2}$, where
        $E$ is the energy, from linear Poisson--like behavior to a saturation.
Berry's computations were not completely
        rigorous, and he also used
        tacitly some assumptions on nondegeneracy of the
        spectrum.
In the present work we carry out a  rigorous study of
        the satistics for all cases in which $S\to\infty $
as $E^\de $ for $\de\ge 1/2$.

\beginsection  Informal Statement of Results \par

        We will consider averages related to the classical
        limit theorems of probability theory.
        Let $N(E,S)$ be the number of $E_n$ in $(E,E+S]$. The question is,
        what is the asymptotics of $\Var N(E,S)$ and what is the limit
        distribution of $\di{N(E,S)-\langle N(E,S)\rangle\over {\sqrt {\Var
N(E,
S)}}}$?
        This problem  has an obvious interpretation
        as a lattice problem.
Referring to (1.1),
        $$
        N(E,S)=\# \{ n\, | \,
        E <
I(n_1-\a_1,n_2-\a_2)
\le E+S\,\}
        $$
 is the number of lattice points in the annulus
        $$
        E < I(x_1-\a_1,x_2-\a_2) \le E+S\,.
        $$
        What we prove in the present article can be briefly summarized as
        follows. Let
        $$
        N(E)=\#\{E_n\: E_n\le E\}.
        $$ Then obviously
        $$
        N(E,S)=N(E+S)-N(E).
        $$
        We prove that if $E\to\infty$, $S/E\to 0$ and $S/E^{1/2}\to\infty$,
then
        $N(E+S)$ and $N(E)$ are asymptotically independent, so
        $$
        \Var N(E,S)\sim\Var N(E+S)+\Var N(E).
        $$
        It was shown in [Bl2] that
        $$
        \Var N(E)\sim V_0E^{1/2},
        $$
        hence
        $$
        \Var N(E,S)\sim 2V_0E^{1/2}.
        $$
        Similarly, the distribution of
        $$
        {N(E,S)-\langle N(E,S)\rangle \over E^{1/4}}
        $$
        converges to the distribution of a difference of two
        independent identically distributed random variables,
        whose distribution coincide with the limit distribution
        of $F(E)=(N(E)-\langle E\rangle )/E^{1/4}$. The existence of a limit
        distribution of $F(E)$ for the circle problem were proved
before in [H-B] and [BCDL]. Results for general ovals were
 proved in [Bl2], and properties
        of this limit distribution were studied in [Bl3]. It
        was shown that in a generic case this limit distribution
        possesses a density which decays at infinity roughly
        as $\exp(-Cx^4)$.
(The distribution $F(E)$ and parameters $V_0,C$, etc.,
 of course depend on $I$ and on $\a$ but we do not indicate this
explicitly).

        In the regime  $E\to\infty$
        and $S/E^{1/2}\to z>0$, we prove a scaling behavior
        of the variance,
        $$
        \Var N(E,S)\sim E^{1/2}V(E^{-1/2}S),
        $$
        and we compute the scaling function $V(z)$ as an
        infinite series. This gives  $V(z)$ as
        an almost periodic function, so it is oscillating
        and has no limit at infinity. We show that
        $$
        \liT\IO V(z)\,dz=2V_0,
$$
        and in a generic case (for instance, for almost all ellipses),
        $$
        V(z)\sim z,\eqno (1.2)
        $$
        as $z\to 0$. This implies that when $z\to 0$,
        $$
        \Var N(E,S)\sim S, \eqno (1.3)
        $$
        which is consistent with a  Poisson
        distribution. The relation (1.2) is violated in
        degenerate cases. For instance, we show that
        for the circle with center at every rational point
 $\a=(\a_1,\a_2)$  the behavior of $V(z)$ is given by
        $$
        V(z)\sim Cz|\log z|,\quad z\to 0. \eqno (1.4)
        $$
This anomalous behavior of $V(z)$ is related to an arithmetic degeneracy
of the circle problem: for some $k\in\N$ there are many representations of $k$
as sum of two squares, so that there are many lattice points at
the circle $\{|x|=k^{1/2}\}.$ On the average, the number of representations
grow
   s
as $\log k$, which shows up in the log--correction to linear
asymptotics of $V(z)$ as $z\to 0$. It is to be noted that recently
Luo and Sarnak [LS] found deviations from the Wigner statistics
for an ``arithmetic'' hyperbolic system (see also earlier physical paper
[BGGS]), which are related as well to an arithmetic degeneracy of the problem.
For a circle with center  at a Diophantine  point $\a $ in $[0,1]^2$,
 the behavior is normal, satisfying (1.3).

        We prove also the existence of a limit distribution
        of
        $$
        {N(E,S)-\langle N(E,S)\rangle\over \sqrt
        {\Var N(E,S)}}
        $$
        in the regime $S/T\to z$. The limit distribution
        is not Gaussian and in a generic case its density
        decays at infinity roughly as $\exp(-Cx^4).$
        However when $z\to 0$ this limit
        distribution converges to
        a standard Gaussian distribution, which can be taken as
        signalling an approach to a random regime.

\beginsection  Precise Formulation of Problem and Results \par

Let $I(x)$ be a homogeneous convex function of order $2$ on the plane, so that,
$$
I(\la x )=\la ^2 I(x)>0,\quad \forall\,\la >0,\; x\in \R^2\setminus \{0\},
\eqno
    (1.5)
$$
$$
\left ( {\partial ^2 I(x) \over {\partial x_i\partial x_j}} \right )_{i,j=1,2}>
0,\quad \forall\,x\in\R^2\setminus \{ 0\}. \eqno (1.6)
$$
We will assume in addition that
$$
I(x)\in C^7(\R^2\setminus \{0\}). \eqno(1.7)
$$
Let $\a\in \R^2.$ Consider
$$
N_0(E;\a)=\#\,\{n\in\Z^2\,:\, I(n-\a)\le \, E\},
$$
which gives the number of lattice points in the convex region $\{x\in\R^2\:
I(x-
   \a)\le E\}.$
We are interested in the behavior of $N_0(E;\a)$
when $\a $ is fixed and $E\to \infty $.

In what follows we will use the parameter $R=E^{1/2}$ instead of $E$,
and we define
$$
N(R;\a)=N_0(R^2;\a).
$$
Then $N(R;\a)$ has a geometric interpretation as the number of lattice points
in
   side the
convex oval $\a +R\g$, where $\g=\{x\in\R^2\: I(x)=1 \}$.
For large $R$, $N(R;\a)$ is approximately equal to the area of
the interior of $\a +R\g$, which is
$$
\text {Area \{ Int }(\a+R\g)\}=\,AR^2,\quad A=\text { Area \{ Int }\g\},
\eqno (1.8)
$$
so the problem is the behavior of the error function
$$
 \De N (R;\a)=N(R;\a)-AR^2 . \eqno(1.9)
$$

Fig. 1 presents $\De N (R;\a)$ for the circle centered at $\a=0$, so that this
i
   s the
error function of the classical circle problem.
 $\De N  (R;\a)$ behaves very irregularly also for other $\a $,
so we may
think of $\De N (R;\a)$ as a random
function of $R$, and ask, what are the statistical properties of $\De N
(R;\a)$?
By statistical properties we mean some averages of functions
of $\De N(R;\a)$,
obtained by weighing $R$ according to some weight.
The statistical properties of $\De N (R;\a)$ were studied
in [H--B] and [BCDL] for a circle and in [Bl2], [Bl3]
for a general oval curve of the following class which
includes (1.5)--(1.7):

{\bf Class $\G$.} {\it $\g\in \G$ if $\g$ is a $C^7$-smooth convex closed
curve without selfintersection in a plane, such that the origin lies inside
$\g$
    and
the curvature of $\g$ is positive at every $x\in\g$.}

Let us normalize $\De N (R;\a) $ to
$$
\FR=R^{-1/2}\ER, \eqno (1.10)
$$
and denote
by  $\Cb$ the space of bounded continuous functions on $\R^1$.
Let$$
0\le c_1<c_2 \eqno (1.11)
$$
be fixed numbers, and $\f (c)\ge 0$ be a fixed bounded
density on $[c_1,c_2]$ with normalization
$$
{1\over {c_2-c_1}}\int_{c_1}^{c_2} \f(c) dc =1, \eqno (1.12)
$$

{\bf Theorem A}  (see [Bl2]). {\it Assume $\g\in \Gamma$. Then there exists
a probability measure $\nu_\a(dt)$
on $\R^1$ such that $\forall\, g(t)\in \Cb$ and $\forall \f(c)$,
$$
\liT\IC g(\FR)\dR
=\int_{-\infty} ^{\infty}g(t)\nu_\a (dt). \eqno (1.13)
$$
In addition,
$$
\liT\IC \FR\dR
=\int_{-\infty} ^{\infty} t \,\nu_\a (dt)=0,
\eqno (1.14)
$$
and
$$
\liT\IC \FR^2\dR
=\int_{-\infty} ^{\infty}t^2\nu_\a (dt).
\eqno (1.15)
$$ }

Note that $\nu_\a(dt) $ does not depend on $\f(c)$.
$\f(c)\equiv 1$ corresponds to a uniform distribution
of $R$ on $[c_1T,c_2T]$, and $\f(c)=2(c_1+c_2)^{-1}c$ corresponds to
a uniform distribution of $E=R^2$ on $[(c_1T)^2,(c_2T)^2]$.

Theorem A shows that typical values of $\De N (R;\a)$ are of order $R^{1/2}$,
and $\nu_\a(dt)$ is a limit distribution of $R^{-1/2}\De N(R;\a)$
assuming that $R$ is nicely distributed on $[c_1T,c_2T]$ and
$T\to \infty$.

In the present work we are interested in the further
investigation of statistical properties of $\ER$. More precisely,
we are interested in the statistics of the increment
$\De N (R+w;\a)-\La(R;\a)$. This increment has a clear geometric
meaning as a difference between the number of lattice points
in the annular strip $\Pi (R,w;\a)$
 between two ovals, $\a+R\g$ and $\a+(R+w)\g $, and
the area of $\Pi(R,w;\a)$.
Our aim is to find  the statistics of $\La(R+w;\a)-\La(R;\a)$.
To formulate the problem precisely
we fix the area $S$ of $\Pi(R,w;\a)$.
We will assume, for  normalization, that
$$
\text {Area \{ Int }\g \} = 1 \eqno(1.16)
$$
Then $w>0 $ is a positive solution of the quadratic equation
$$
2w R +w^2 =S, \eqno(1.17)
$$
$$w \, =\,(S/R)\, (1+(1+(S/R^2))^{1/2})^{-1}=(S/(2R))\,
(1+O((S/R^2)^2)) \eqno (1.18)
$$
Let
$$
\La(R,S;\a)=\La(R+w;\a)\,-\,\La(R;\a)
\eqno (1.19)
$$
with $w$ given by (1.18). Then
$$
\La(R,S;\a)=N_0(E+S;\a)-N_0(E;\a)-S,\quad E=R^2.
$$
Define
$$
D(T,S;\a)=
{1\over {(c_2-c_1) T}} \int_{c_1T}^{c_2T} (\La(R,S;\a))^2\dR. \eqno (1.20)
$$

{\bf Scaling Behavior}
$$\eqalignno{\text{(I)}\quad&\lim_{S/T^2\to 0,\, S/T\to\infty} T^{-1}D(T,S;\a)
=V(\a)>0;&(1.21)\cr
\text{(II)}\quad&\lim_{T\to\infty,\,  S/T\to z} T^{-1}D(T,S;\a)=V(z;\a)>0,\,
\fo
   rall
\, z>0;&(1.22)\cr
\text{(III)}\quad&\lim_{z\to 0} z^{-1}V(z;\a)=1,
\text { for typical } \g .&(1.23)\cr}
$$

We prove  that (I), (II) hold for all $\g\in\G$
and compute $V(\a )$ and $V(z;\a)$.
(III)  will
be shown to hold for generic $\g$, i.e.,
when $\g$ has no symmetries which give rise
to multiplicities of the eigenvalues.
In general, the behavior of $V(z;\a)$,
as $z\to 0$ will depend on the relevant group of symmetry.

For a Poisson point random field of density $1$ in the  plane
the variance of the number of points in a domain of area $S$ is equal to $S$,
so

Eqs. (1.21)--(1.23)  describe a transition from a Poisson--like asymptotics
at (III) (for typical $\g $) through a scaling
at (II) to a saturation at (I).

{\it Notation }.
For $\xi\in\R^2,$ $\xi\not = 0$, consider a point $x(\xi )\in \g $,
 where the outer normal vector to $\g $, $n_{x(\xi)}$,   coincides
with $|\xi|^{-1}\xi$.
Denote by $\r (\xi) $ the radius of curvature of $\g $ at $x(\xi )$
and by
$$
Y(\xi )=\xi\cdot x(\xi).\eqno(1.24)
$$
where $a\cdot b=a_1b_1+a_2b_2$ for $a,b\in\R^2$.
The curve $\g^*=\,\{\xi:\, Y(\xi )=1\,\}$
is known as the polar of $\g$.
Note that $(1/2)Y^2 (\xi)$ is the Legendre transform of $(1/2)I(x)$,
$$
(1/2)Y^2(\xi)=-\inf _{x\in\R^2} ((1/2)I(x)-x\cdot \xi). \eqno(1.25)
$$

Let $0<Y_1<Y_2<\dots $ be all possible values of $Y(n)$
with $n\in \Z^2\backslash  \,\{ 0\}$.
Define
$$
u_\a (k)=\sum_{n\in\Z^2\,:\,Y(n)=Y_k}e(n\cdot \a)|n|^{-3/2}
\sqrt {\r(n)}, \eqno (1.26)
$$
where
$$
e(t)=\exp (2\pi it).\eqno (1.27)
$$
Let
$$
J_\f=\Ic c\,\f(c)\,dc.\eqno (1.28)
$$

        {\bf Theorem 1.1} {\it Assume $\g\in\G$. Then
        $$
        \lLT T^{-1}\,\IC(\ERL)^2\dR
=V(\a)=J_\f W(\a)
        \eqno (1.29)
        $$
        with
        $$
        W(\a)=\pi^{-2}\Sk |u_\a(k)|^2.
        \eqno (1.30)
        $$}

        {\bf Corollary.} {\it If $\g\in\G$, then
        $$
        \lc\lLT T^{-1}\,\IDC(\ERL)^2dR
        =W(\a).
        \eqno (1.31)
        $$}

        {\bf Theorem 1.2.} {\it If $\g\in\G$, then
        $$\eqalign{
        \lLt T^{-1}\,\IC&(\ERL)^2\dR\cr
        &=V(z;\a)=\Ic c\,\f(c)\,W(z/c;\a)\,dc\cr}
        \eqno (1.32)
        $$
        with
        $$
        W(z;\a)=\pi^{-2}\Sk |u_\a(k)|^2
        (1-\cos(\pi Y_kz)).
        \eqno (1.33)
        $$}

        {\bf Corollary.} {\it If $\g\in\G$, then
        $$
        \lc\lLt T^{-1}\,\IDC(\ERL)^2dR
        =W(z;\a).
        \eqno (1.34)
        $$}

The scaling function  $W(z;\a)$ represent a local averaging of $(\ERL)^2$.
        Comparing $(1.31)$ with $(1.34)$ one would expect that
        $\lim_{z\to\infty}W(z;\a)=W(\a)$. Formula (1.33) shows,
        however, that this is not true:
        $W(z;\a)$ is an almost periodic function of  $z$, so it
        oscillates at infinity. What is true is that on the average
        $W(z;\a)$ converges to $W(\a)$, that is
        $$
        \liT\IO W(z;\a)\,dz=W(\a).\eqno (1.35)
        $$

        An important problem is the behavior of $W(z;\a)$ when
        $z\to 0$. We shall prove that it is universal in a generic
        situation. To that end we introduce the following class
        of ovals:

        {\bf Class $\G_0$.} {\it $\g\in\G_0$, if $\g\in\G$
        and $Y(m)=Y(n)$ only if $m=n$.}

Since $\G_0$ is defined through a countable number of inequalities
\{$Y(m)\not= Y(n)$, $m\not= n$\}, we may think of $\G_0$ as of the set
of generic ovals.
        In the case when the oval $\g$ is symmetric
        with respect to the origin, the
        condition that $Y(m)=Y(n)$ only if $m=n$
        is certainly violated, since then $Y(-n)=Y(n)$.

        To cover this case we introduce the following
        class of ovals. Let $G$ be the group of all isometries
$i\:\R^2\to\R^2$ of a plane such that $i(0)=0$ and $i(\Z^2)=\Z^2$.
$G$ consists of 8 elements.
Let $H$ be a subgroup of $G$. We say that an oval $\g$ is invariant
with respect to $H$ if $g\g=\g$ for every $g\in H$.

        {\bf Class $\G_0(H)$.} {\it $\g\in\G_0(H)$, if $\g\in\G$
        is invariant with respect to $H$,
        and $Y(m)=Y(n)$ only if $m=gn$ for some $g\in H$.}

For $\g\in\G_0(H)$ and $\a\in\R^2$, consider a subgroup $H_\a\subset H$
such that $g\in H_\a$ if $g\a=\a+n$ with some $n\in\Z^2$. Let
$$
m_\a(H)=|H_\a|
$$
be the number of elements in $H_\a$.

        {\bf Theorem 1.3.} {\it If $\g\in\G_0(H)$ then
        $$
        \lim_{z\to 0}z^{-1}W(z;\a)=m_\a(H).
        \eqno (1.37)
        $$}

{\it Remarks.} 1. If $\g\in\G_0$, (1.37) reduces to
$ \lim_{z\to 0}z^{-1}W(z;\a)=1.  $
2. Note, that if $m\in\a+R\g$ and $g\in H_\a$,
so that $g\a=\a+n$, then $gm\in g\a+Rg\g=\a+n+R\g$, hence $gm-n\in \a+R\g$.
For a typical $m\in\Z^2$ the points $gm-n$, $g\in H_\a$,
are different, therefore
$m_\a(H)$ is the multiplicity of integer points on
$\a+R\g$, caused by the symmetry.

        A circle with center at the origin  is invariant with respect to
$G$. However the circle does not belong to $\G_0(G)$, since
for some $k\in\N$
there exist  many different representations of $k$ as a sum of
two squares, which are not related to any symmetry. The number of such
represent
   ations
grows, on the average, as $\log k$ and this shows up in the behavior of
$W(z;0)$
   .

A vector $\a\in\R^2$ is called Diophantine if there exist $C,N>0$
such that for all $n\in\Z^2\setminus\{0\}$,
$$
|n\cdot \a|\ge C|n|^{-N}.
$$

        {\bf Theorem 1.4.} {\it If $\g$ is a circle with the center
        at the origin, then
        $$\eqalign{
        \lim_{z\to 0}(z|\log z|)^{-1}W(z;\a)
        &=C_\a>0 \quad\text
        {for every rational $\a\in\Q^2$},\cr
        \lim_{z\to 0}z^{-1}W(z;\a)
        &=1\quad \text{for every Diophantine $\a$}.\cr}
        $$}

{\it Remarks.} 1. From the proof of Theorem 1.4 below an explicit
formula for $C_\a$ follows. For instance, for $\a=0$,
$C_\a=6\pi^{-1}$.

2. Theorem 1.4 can be extended to any
ellipse with rational ratio
of squared half--axes $a_1^2/a_2^2$.
We are going to consider this problem elsewhere.

Theorems 1.3, 1.4 are illustrated in Figs. 2--4, which
present the scaling function $W(z;\a)$, respectively,
for a circle with $\a=(0,0)$, for an ellipse with ratio of axes $1/\pi$
and $\a=(0,0)$, and finally for the same ellipse with
$\a=(0.1,0.1)$.
Different behavior of $W(z;\a)$ as $z\to 0$ is well seen
in the figures as predicted in Theorems 1.3, 1.4.

        Now we consider the  existence
        of a limit distribution of $ F(R,S;\a)=R^{-1/2}\La (R,S;\a).$
Let  $\nu_\a(dt)$ be the limit distribution
        of $F(R;\a)$ (see theorem A above).
        Denote by $\nu_\a^-(dt)$ the distribution obtained by reflection of
 $\nu_\a(dt)$,
        so that
        $$
        \int_a^b\nu_\a^-(dt)=\int_{-b}^{-a}\nu_\a(dt).
        $$

        {\bf Theorem 1.5.} {\it If $\g\in\G$ then $\forall\; g(t)\in\Cb$,
        $$
        \lLT\IC g(\FRL)\,\f(R/T)\,dR=\II g(t)\,\mu_{\a}(dt)
        $$
        with $\mu_\a=\nu_\a *\nu_\a^-$, where $*$ denotes convolution.}

        A simple explanation of Theorem 1.5 is that when
        $S/T\to\infty$, i.e., when the annulus is ``thick'',
        $F(R+w;\a)$ and $F(R;\a)$ are independent
        random variables in the limit $T\to\infty$,
        assuming that $R$ is uniformly distributed on $[c_1T,c_2T]$,
        so $$
        F(R,S;\a)\approx F(R+w;\a)-F(R;\a)
        $$
        is a difference of independent random variables.

        {\bf Theorem 1.6.} {\it If $\g\in\G$ then for
        every $z>0$ there exists a probability measure $\mu_\a(dt;z)$,
which depends continuously, in the weak topology, on $z$,
        such that $\forall\; g(t)\in \Cb$,
        $$\eqalign{
        \lLt&\IC g(F(R,S;\a))\,\dR\cr
&=\Ic dc\,\f(c)\II g(t)\,\mu_\a(dt;z/c).\cr}
        $$}

{\bf Corollary.}
        $$
        \lim_{\Delta c\to 0}
        \lLt{1\over T\Delta c}\int_T^{T(1+\Delta c)}
        g(F(R,S;\a))\, dR=\II g(t)\,\mu_\a(dt;z),
        $$

        {\bf Theorem 1.7.} {\it For every
$g(t)\in\Cb$,
        $\II g(t)\mu_\a (dt;z)$ is a continuous
        almost periodic function in $z$, and
        $$
        \liT\IO dz\II g(t)\mu_\a (dt;z)=\II g(t)\mu_\a(dt).
        $$}

Our next goal is to describe properties of the measures
$\matt$.
To do this we need some conditions of
incommensurability of the frequencies
$Y(n)$. Define $M\subset \Z^2$ as $M=M_0\cup M_1$ with
$$\eqalign{
&M_0=\{(0,1),(1,0),(0,-1),(-1,0)\},\cr
&M_1=\{n=(n_1,n_2)\mid n_1n_2\not=0;\;|n_1|,|n_2|
\quad\text{are positive and relatively prime}\}.\cr}
$$
In some respects  $M$ plays the role of the set
of prime numbers on the lattice $\Z^2$. Define now

{\bf Class $\G_1$.} {\it An oval $\g$ belongs to $\G_1$, if
$\g\in\G$ and the numbers $\{Y(n);\;n\in M\}$ are linearly
independent over $\Q$.}

It is clear that $\G_1\subset\G_0$, so $\G_1$ does not
contain symmetric ovals. To cover the case of symmetric
ovals let us consider a subgroup $H\subset G$.
Consider any fundamental domain $M(H)\subset M$ for
the action of $H$ on $M$, i.e., for every $n\in M$
there exists a unique $m\in M(H)$ such that $n=gm$
for some $g\in H$. Define

{\bf Class $\G_1(H)$.}
{\it An oval $\g$ belongs to $\G_1(H)$
if $\g$ is invariant with respect to $H$,
and the numbers $\{Y(n);\;n\in M(H)\}$ are linearly
independent over $\Q$.}

As an example consider
$$
g_0\:(x_1,x_2)\to -(x_1,x_2),\quad g_1\:(x_1,x_2)\to (-x_1,x_2),
\quad g_2\: (x_1,x_2)\to (x_1,-x_2),
$$
and
$$
\eqalign{
H_0&=\{\Id ,\;g_0\},\cr
H_{12}&=\{\Id ,\;g_0,\;g_1,\;g_2\},\cr}
$$
which are, respectively, the   symmetry group with respect to the
origin and the  symmetry group with respect to the
coordinate axes.
A general class of ovals which belong to $\G_1(H_0)$ and $\G_1(H_{12})$
is described in [Bl3]. A characteristic example is an
ellipse with transcendental ratio of half--axes. In
general, the condition that the numbers $Y(n),\; n\in M(H)$,
are linearly independent over $\Q$ (or, equivalently, over $\Z$),
is equivalent to a countable number of inequalities
$$
\sum_{k=1}^N r_kY(n_k)\not=0,\quad r_k\in\Z,
$$
and this condition can be viewed as a condition of
a ``generic'' situation.

{\bf Theorem 1.8.} {\it If $\g\in\G_1(H)$ for some $H\subset G$,
then for every $z>0$, $\matt$ possesses a density
$$
\ptz={\matt\over dt},
$$
which is an analytic (entire) function of $t\in\C$, and
for real $t$,
$$
\eqalignno{
0\le \ptz&\le C\exp(-\la t^4),&(1.38)\cr
P_\a(-t;z),\;1-P_\a(t;z)&\ge C'\exp(-\la' t^{4}),
\quad t\ge 0,
&(1.39)}
$$
where $P_\a(t;z)=\di\int_{-\infty}^t p_\a(t';z)\,dt'$
and $C,\la,C',\la'>0$.}

{\bf Theorem 1.9.} {\it The previous theorem holds for
a circle with the center at the origin, with a slightly weaker
estimates, instead of (1.38), (1.39):
$$\eqalign{
0\le\ptz
&\le C(\ep)\exp(-\la(\ep)t^{4-\ep}),\cr
P_\a(-t;z),\;1-P_\a(t;z)&\ge C_1(\ep)\exp(-\la(\ep)t^{4+\ep}),\quad
\forall\;\ep>0,\quad C(\ep),\la(\ep),C_1(\ep)>0,\cr}
\eqno (1.40)
$$}

If $\mu(dt)$ is a distribution of a random variable $\xi$
with zero mean, denote by $\tilde \mu(dt)$ a distribution of the
normalized random variable $\xi/\sqrt{\Var \xi}$. Then
$$
\II t^2 \tilde \mu(dt)=1.
$$
Now we describe the limit behavior
of the measure $\tilde \mu_\a(dt;z)$ when $z\to 0$.

        {\bf Theorem 1.10.} {\it If $\g\in\G_1(H)$ for some $H\subset G$,
then $\lim_{z\to 0}\tilde\mu_\a(dt;z)$ is a standard Gaussian distribution.}

        {\bf Theorem 1.11.} {\it If $\g$ is a circle with the center
        at the origin then $\lim_{z\to 0}\tilde \matt$ is a standard Gaussian
di
stribut
   ion.}

The proof of the above results makes use of the fact,
first noted by Heath-Brown [HB], that the parameter $R$ in $F(R;\a)$
can be thought of as a time parameter
in a flow on an infinite dimensional torus.
Statistical properties of $F(R;\a)$ are therefore related
to ergodic properties of almost
periodic functions. These in turn can be obtained by suitable approximations
as quasiperiodic functions, i.e., by flows on a finite dimensional torus,
which is a part of standard ergodic theory.
To carry out this program we should devote the next section
to the derivation of some general result on the ergodic properties
of almost periodic functions in the Besicovitch space $B^2$. Before doing that,
however, we restate a theorem from [Bl2] which shows that $F(R;\a)$ belong to
$B
   ^2$.

        {\bf Theorem B. } {\it If $\g \in \Gamma$, then
        for every $\a\in\R^2$, $ \FR $, as a function
        of $R$, belongs to the Besicovitch space $B^2$.
        A Fourier expansion of $\FR$ in $B^2$  is given by the formula
(see Sect. $ 2$)
        $$
        \FR
        =\PI\SZ\NS\CS
        \eqno (1.41)
        $$
        with
        $$
        \phi(n;\a)=2\pi n\cdot \a-\tf.
        $$}

Using the notation (1.26) we can rewrite the Fourier series (1.41) as
$$
\FR=\PI\Sk |u_\a(k)|\,\cos(2\pi Y_kR+\theta_\a(k)-\tf),
\quad \theta_\a(k)=\arg u_\a(k).\eqno (1.42)
$$

We shall derive Theorems 1.1, 1.2, and 1.5--1.7 from Theorem B and some general
results on almost periodic functions in $B^2$. These
general results
are formulated and proved in the next section. To prove
Theorems 1.3, 1.4 and 1.8--1.11 we need more refined arguments.

        \beginsection 2. Some general results on almost periodic
 functions \par

        In this section we will prove some preparatory results
on a limit distribution of the values of an almost periodic function.
        We will use the
        Besicovitch space $B^2$ of almost periodic functions.
        A function $F(R)$
        on $\{0<R<\infty\}$ belongs to $B^2$ if for every
        $\ep>0$ there exists a trigonometric polynomial
        $$
        P_\ep(R)=\sum_{n=1}^{N_\ep}a_{n\ep}\exp(i\la_{n\ep}R)
        \eqno (2.1)
        $$
        such that
        $$
        \lT\IO|F(R)-P_\ep(R)|^2\,dR<\ep \eqno (2.2)
        $$
        For $F(R)\in B^2$ we can define (see [Bes])
        $$
        \|F(R)\|_{B^2}=\left(\lim_{T\to\infty}\IO|F(R)|^2dR\right)^{1/2}
        \eqno (2.3)
        $$
        It is to be noted that $\|\cdot\|_{B^2}$
        is only a seminorm and not a norm,
        i.e., $\|F(R)\|_{B^2}=0$ does not imply $F(R)\equiv 0$.
        For instance, if $\lim_{R\to\infty}F(R)=0$,
        then $\|F(R)\|_{B^2}=0$. The Fourier
        coefficients of $F(R)\in B^2$ are defined as
        $$
        a(\la)=
        \lim_{T\to\infty}\IO F(R)\exp(-i\la R)\,dR
        \eqno (2.4)
        $$
        It is known that $a(\la)\not=0$ at most only for countably many
        $\la=\la_n$, $n\in\N$, and that $\|F(R)\|_{B^2}=0$ iff all
        $a(\la_n)=0$.
        We shall use the notation
        $$
        F(R)=\sum_{n=1}^\infty a(\la_n)\exp(i\la_nR),
        \eqno (2.5)
        $$
        which shows that $a(\la_n)$ are the Fourier coefficients
        of $F(R)$.

        For $F(R)\in B^2$ we have the Parseval identity (see [Bes]):
        $$
        \lim_{T\to\infty}\IO |F(R)|^2dR
        =\sum_{n=1}^\infty |a(\la_n)|^2, \eqno (2.6)
        $$
and
$$
\lN\lT\IO|F(R)-\sum_{n=1}^N a(\la_n)\exp(i\la_nR)|^2dR=0.
\eqno (2.7)
$$
        If $F(R)$ is real--valued, then $a(-\la_n)=\overline {a(\la_n)}$,
and the Fourier series (2.5) can be rewritten as
        $$
        F(R)=\sum_{n=1}^\infty b(\la_n)\cos(\la_nR+\phi_n).
\eqno (2.8)
        $$
        Then the Parseval identity has the form
        $$
        (\| F(R)\|_{B^2})^2=b(0)^2+(1/2)\sum_{n\:\la_n\not=0}
        b(\la_n)^2.\eqno (2.9)
        $$
        We have also a more general formula:
        $$
        \liT\IO F(R+t)F(R)\,dR
        =b(0)^2+(1/2)\sum_{n\:\la_n\not=0}b(\la_n)^2
        \cos(\la_nt).
\eqno (2.10)
        $$
        The Schwarz inequality and (2.2) imply
        $$
        \lT\IO\min\{1,|F(R)-P_\ep(R)|\}\,dR<\ep.\eqno (2.11)
        $$

        We shall use the following theorem from [Bl2]:

        {\bf Theorem C.} {\it If $F(R)\in B^2$, then there exists a
        probability distribution $\nu (dx)$ on $\R^1$,
        with a finite variance
        $\II x^2\nu(dx)$, such that for every
        probability density $\f (x)$ on $[0,1]$ and every
        bounded continuous function $g(x)$ on $\R^1$,
        $$
        \liT  \IO g(F(R))\f (R/T)dR=\II g(x)\nu (dx). \eqno (2.12)
        $$
        In addition,
        $$
        \liT\IO F(R)\,dR=\II x\nu(dx)=a(0)
        $$
        and
        $$
\liT\IO |F(R)|^2\,dR=\II x^2\nu(dx).
        $$ }

        The distribution $\nu(dx)$ defined by (2.12) is called the
        distribution of $F(R)$.

        The definition of the space $B^2$ and all the discussed
properties
        of almost periodic functions from the space $B^2$, including
        Theorem C, admit a straightforward extension to
        vector--valued functions
        $F(R)=(F_1(R),$ $\dots F_k(R))$.

        We shall call a joint distribution of $k$ almost periodic
        functions $F_1(R),\dots,F_k(R)$ the distribution of the
        vector--function $F(R)=(F_1(R),\dots,F_k(R))$.
        Here, in principle, $F_j(x)$ can also be a vector--valued
        almost periodic
        functions, but we shall not use this case.

        It is noteworthy that if the Fourier frequencies
        $\{\la_n^{(1)}\},\dots,\{\la_n^{(k)}\}$
        of $F_1(R),\dots, F_k(R)$ are linearly independent
        over $\Q$, i.e., if
        $$
        \sum_{n=1}^{N_1} r^{(1)}_n\la^{(1)}_n
        +\dots+
        \sum_{n=1}^{N_k} r^{(k)}_n\la^{(k)}_n
        =0,
        \quad r^{(i)}_n\in\Q,
        $$
        implies
        $$
        \sum_{n=1}^{N_1} r^{(1)}_n\la^{(1)}_n
        =\dots=
        \sum_{n=1}^{N_k} r^{(k)}_n\la^{(k)}_n=0,
        $$
        then a joint distribution of $F_1(R)\dots,F_k(R)$
        is a product of the distributions of $F_1(R)\dots,F_k(R)$, so that
        $F_1(R)\dots,F_k(R)$ are independent.

        If $G$ is a region in $\R^k$, we shall, as usual, call
        $L^\infty(G)$ the space of measurable bounded functions
        on $G$. As before, we will assume that $\f(c)\ge 0$ is a
function from $L^\infty([c_1,c_2])$, which satisfies (1.7).

        We shall prove the following theorem:

        {\bf Theorem 2.1. } {\it Let $F(R)\in B^2$,
        $w=\wL>0$ be a positive solution of equation (1.17), and
        $g(x)=g(x_1,x_2)$ be a continuous function on $\R^2 $,
        such that $g(x_1,x_2)=O(x_1^2+x_2^2)$, when $x_1^2+x_2^2\to\infty$.
Then
        $$\eqalign{
        \lLT
        &\IC g(F(R+\wL),F(R)) \,\f(R/T)\,dR\cr
        &=
        \II \II g(x_1,x_2) \nu (dx_1) \nu (dx_2),\cr}
         \eqno (2.13)
        $$
        where $\nu(dx)$ is the distribution of $F(R)$.}

        The condition $S/T\to\infty$ implies that $w_S(T)\to\infty$.
        Theorem 2.1 shows that in this case,
        assuming also that $S/T^2\to 0$,
        $F(R+\wL))$ and $F(R)$ are
        asymptotically independent. A heuristic explanation of this result is
that the averaging in two different scales, $R$ and
        $\wL$, is independent in the limit $T\to \infty $.

        {\it Proof of Theorem 2.1.}
        We will prove (2.13) first in a particular case, when
        $g(x)=g(x_1,x_2)\in C_0^\infty (\R^2)$, i.e., $g(x)$
is a $C^\infty$--function with compact support,
        and $\f(c)\equiv 1$.
        We will  then consider the general case.

        The condition $g(x)\in C_0^\infty(\R^2)$ implies that
        for all $x,y \in \R^2$,
        $$
        |g(x)-g(y)|\le C_0 \min \{ 1,\ |x-y| \, \} \eqno (2.14)
        $$
        with some $C_0>0$. So from (2.11),
        $$
         \IC |g(F(R+\wL),F(R)) -
        g(P_\ep (R+\wL),P_\ep (R))|\,dR \le C_1\ep \eqno(2.15)
        $$
        with some $C_1>0$.

        We can choose frequencies $\om _1,\dots , \om _k $, which are
linearly independent over $\Q $,
        such that all $\la _{n\ep }$ in (2.1) are linear
        combinations of $\om _1,\dots , \om _k $,
        with integer coefficients. Then
        $$
        P_\ep (R)=\sum _{n\in M} a_{n\ep } \exp (in\om R)  \eqno(2.16)
        $$
        where $M \subset \Z^k $ is a finite set of multiindices
        $n=(n_1,\dots , n_k)$ and
        $n\om =n_1\om _1+ \dots + n_k \om _k $. Define
        $$
        A_\ep (t_1, \dots , t_k)=\sum _{n\in M} a_{n\ep } \exp (int)
        \eqno(2.17)
        $$
        with $t=(t_1,\dots , t_k)$ and
        $nt =n_1t _1+ \dots + n_k t _k $.
        Then
        $$
        P_\ep (R) = A_\ep (\om_1 R,\dots , \om_k R) , \eqno (2.18)
        $$
        hence
        $$ \eqalign {
        &\IC
        g(P_\ep
        (R+\wL),P_\ep (R))dR\cr
        &=\IC g(
        A_\ep (\om _1(R+\wL),\dots ,\om _k(R+\wL)),
        A_\ep (\om _1R,\dots ,\om _kR )) dR.\cr}
        $$
        We will use the following general formula: If $f(R)$ is a bounded
        continuous function on $[0,\infty )$, then for $U,T\to \infty $
        with $U/T\to 0 $,
        $$
        \IB f(R)dR=\IB {1\over U} \int_R^{R+U} f(Q)dQ dR + O(U/T).
\eqno(2.19)
        $$
Indeed, the integral on the RHS is equal to
$$\eqalign{
&{1\over (c_2-c_1)TU}\int\int_{\{c_1T\le R\le c_2T,\;0\le Q-R\le
U\}}f(Q)\,dQdR\
   cr
=&{1\over (c_2-c_1)TU}\int\int_{\{c_1T\le Q\le c_2T,\;0\le Q-R\le
U\}}f(Q)\,dQdR
+O(U/T)\cr
=&{1\over (c_2-c_1)T}\int_{\{c_1T\le Q\le c_2T\}} f(Q)\,dQ+O(U/T), \cr}
$$
which coincides with the LHS up to $O(U/T)$.
(2.19) implies that
        $$\eqalign{
        \IB
        &g(P_\ep (R+\wL),P_\ep (R))dR\cr
        &=\IB {1\over U}\int_R^{R+U} g(P_\ep(Q+w_S(Q)),P_\ep(Q))dQdR
        +O(U/T).\cr}
        \eqno (2.20)
        $$
 From (1.18),
        $$
        |w_S(R)-w_S(Q)|\le C|R-Q|ST^{-2}\le CUST^{-2},
\eqno (2.21)
        $$
        when $ST^{-2}\ll 1$, $R\le Q\le R+U$, $0\le U\le T$ and $1\ll c_1T\le
R\
le c_2T
   $, hence
        $$\eqalign{
        \IB&{1\over U}\int_R^{R+U} g(P_\ep(Q+w_S(Q)),P_\ep(Q))dQdR\cr
        &=\IB {1\over U}\int_R^{R+U} g(P_\ep(Q+\wL),P_\ep(Q))dQdR
        +O(ST^{-2}).\cr}
        \eqno (2.22)
        $$
        Using the ergodic theorem (see, e.g., [CFS]), we have
that  for every $\delta>0$,
there is an $U_0(\de)$ such that
        $$ \eqalign {
        & \biggl| {1\over U} \int_R^{R+U}
        g( A_\ep (\om _1(Q+\wL),\dots ,\om _k(Q+\wL)),
        A_\ep (\om _1Q,\dots ,\om _kQ ))dQ \cr
        &-\int_{\T^k}g(A_\ep (t_1+\om_1\wL, \dots ,
        t_k+\om_k\wL),A_\ep(t_1,\dots, t_k))\,dt\biggr|<\delta/2
         \cr }
        \eqno (2.23)
        $$
        when $U>U_0(\delta )$. When this is combined with (2.18)--(2.22),
it yields
        $$\eqalign{
        &\biggl| \IB
        g( P_\ep (R+\wL),
        P_\ep (R))dR\cr
        &-\int_{\T^k}dt\, \IB dR\, g(A_\ep (t_1+\om_1\wL, \dots ,
        t_k+\om_k\wL),A_\ep(t_1,\dots, t_k))\biggr|<\delta\cr}
         \eqno (2.24)
        $$
        whenever $T$ and $S^{-1/2}T$ are sufficiently large.

        We will prove now, that
        $$\eqalign{
        \lLT \IB
        &g(A_\ep (t+\wL\om),A_\ep(t))dR\cr
        &=\int_{\T^k} g(A_\ep (s),A_\ep(t))ds,\cr}
        \eqno (2.25)
        $$
        where $t=(t_1,\dots,t_k)$, $s=(s_1,\dots,s_k)$
        and $\om=(\om_1,\dots,\om_k)$. Moreover, the convergence
in (2.25) is uniform in $t\in\T^k$.

        Define $y=\wL $. From $(1.18)$,
        $$\eqalign{
        y&={S\over 2R}(1+O(S/T^2));\cr
        {dR\over dy}&=-{S\over 2y^2}(1+O(S/T^2)),\cr
        }$$
        when $R\ge c_1T\gg S^{1/2}$, so
        $$\eqalign{
        &\IC g(A_\ep (t+\wL\om),A_\ep(t))dR\cr
        &={1\over (c_2-c_1)T}\int_{w_L(c_2T)}^{w_L(c_1T)}
        g(A_\ep(t+y\om),A_\ep(t))(dR/dy)dy\cr
        &={1\over (c_2-c_1)T}\int_{(2c_2T)^{-1}S}^{(2c_1T)^{-1}S}
        g(A_\ep(t+y\om),A_\ep(t))(S/(2y^2))dy+O(S/T^2).\cr}
        \eqno (2.26)
        $$
        Define $\tau=S/T$ and
        $$
        \f_0(x)={x_1x_2\over x^2(x_2-x_1)}
        $$
        with
        $$
        x_1=(2c_2)^{-1},\quad x_2=(2c_1)^{-1}.
        $$
        Then $\di\int_{x_1}^{x_2}\f_0(x)\,dx=1$ and
        $$\eqalign{
        &{1\over (c_2-c_1)T}\int_{(2c_2T)^{-1}S}^{(2c_1T)^{-1}S}
        g(A_\ep(t+y\om),A_\ep(t))(S/(2y^2))dy\cr
        &={1\over \tau}\int_{x_1\tau}^{x_2\tau}
        g(A_\ep(t+y\om),A_\ep(t))\f_0(y/\tau)\,dy.\cr}
        $$
        By the ergodic theorem the last integral converges
uniformly in $t\in\T^k$, when
        $\tau\to \infty $, to
        $$
        \int_{\T^k} g(A_\ep (s),A_\ep(t))ds,
        $$
        which implies (2.25).

It now follows from (2.24), (2.25) that
        $$\eqalign{
         \lLT
        & \IB g(P_\ep (R+\wL),P_\ep (R))dR\cr
        &= \int_{\T^k }
         \int_{\T^k }
        g( A_\ep (s),
        A_\ep (t)) ds dt.\cr} \eqno(2.27)
        $$
        Let
        $$
        \nu_\ep (B) =\int _{\{ t:A_\ep(t)\in B \} } dt
        $$
        be the distribution of $A_\ep (t)$. Then we can rewrite (2.27) as
        $$\eqalign{
         \lLT
        &\IB g(P_\ep (R+\wL),P_\ep (R))dR\cr
        &= \II \II g(x_1,x_2) \nu_\ep (dx_1)\nu_\ep(dx_2).\cr}
         \eqno (2.28)
        $$
        By the ergodic theorem,
        $$
         \liT \IB g(P_\ep (R))dR=
         \int_{\T^k } g(A_\ep (t))dt=
        \II g(x)\nu_\ep(dx)  \eqno(2.29)
        $$
        for every $g(x)\in C_0^\infty (\R^1)$. By Theorem C,
        $$
        \liT  \IB g(F(R))\,dR=\II g(x)\nu (dx),
        $$
        and by (2.11),
        $$
        \IB |g(F(R))dR-g(P_\ep(R))dR|\le C_0\ep ,
        $$
        so
        $$
        \lim_{\ep\to 0 }\II g(x)\nu_\ep(dx)=\II g(x)\nu(dx).\eqno(2.30)
        $$
        Now (2.28) together with (2.15), (2.30) implies
        $$\eqalign{
        \lLT &
        \IB g(F(R+\wL),F(R)) dR\cr
        &= \II \II g(x_1,x_2) \nu (dx_1) \nu (dx_2).\cr}
        $$
        Hence (2.13) is proved in the particular case when
        $g(x_1,x_2)\in C_0^\infty(\R^2)$
        and $\f(c)\equiv 1$.

        By implication (2.13) holds in the case when
        $g(x_1,x_2)\in C_0^\infty $ and $\f(c)$
        is a step--wise function with a finite number of steps.
Now every $\f(c)\in\LC$ is a limit in $L^1$-norm of
        step--wise functions hence, again by implication,
        (2.13) holds in the case when
        $g(x_1,x_2)\in C_0^\infty(\R^2) $ and $\f(c)\in\LC$.

        Assume now that $g(x_1,x_2)$ is a continuous function with
        compact support. For every $\ep>0$ there exists $g_\ep(x_1,x_2)
        \in C_0^\infty(\R^2)$ such that
        $$
        \sup_{x_1,x_2}|g_\ep(x_1,x_2)-g(x_1,x_2)|<\ep,
        $$
        hence
        $$
        \IB|g_\ep(F(R+\wL),F(R))-g(F(R+\wL),F(R))|\,dR
        <\ep.
        $$
Thus it follows that (2.8) holds for every continuous function $g(x_1,x_2)$
with
        compact support.

        The condition $F(R)\in B^2$ implies that
        $$
        \lim_{A\to\infty}\lT\IB |F(R)|^2\,\chi_{\{|F(R)|\ge A\}}(R)\, dR
        =0.
        $$
        Let $\psi(y)\in C^\infty(\R^1)$ and
        $$
        \psi(y)\quad\eqalign{
        &=0,\quad\text{if}\quad|y|>2A,\cr
        &=1,\quad\text{if}\quad|y|<A,\cr
        &\in [0,1] \quad\text{otherwise}.}
        $$
        Assume that $g(x)$, $x=(x_1,x_2)$, is a continuous function such that
        $g(x)=O(|x|^2)$ as $|x|\to\infty$. Then
        $$\eqalign{
        &\lim_{A\to\infty}\lT\left|\IB
        g(F(R+\wL),F(R))\,(1-\psi(|F(R)|+|F(R+\wL)|))\, dR\right|\cr
        &\le C_0\lim_{A\to\infty}\IB
        (|F(R)|^2\chi_{\{|F(R)|\ge A\}}(R)\cr
        &+|F(R+\wL)|^2\chi_{\{|F(R+\wL)|\ge A\}}(R))\,dR=0.\cr}
        $$
        Hence, by implication,
        (2.13) holds for every continuous function $g(x_1,x_2)$
        with $g(x)=O(|x|^2)$ as $|x|\to\infty$.
        Theorem 2.1 is proved.

        Assume $F(R)\in B^2$. Denote $\nu(dx_1,dx_2;z)$ a distribution
        of the pair $(F(R+z),F(R))$, i.e., for every continuous function
        $g(x_1,x_2)$ on $\R^2$,
        $$
        \liT\IO  g(F(R+z),F(R))\,dR
        =\II\II g(x_1,x_2)\,\nu(dx_1,dx_2;z).
        $$

        {\bf Theorem 2.2.} {\it Under the same assumptions as
        in Theorem 2.1,
        $$\eqalign{
        \lLt\IB
        &g(F(\Rw),F(R))\,\f(R/T)\,dR\cr
        &=\II\II g(x_1,x_2)\Ic dc\,\f(c)\,\nu(dx_1,dx_2;z/(2c)),\cr}
        \eqno (2.31)
        $$
        and
        $$\eqalign{
        \lc\lLt\IDC
        &g(F(\Rw),F(R))\,dR\cr
        &=\II\II g(x_1,x_2)\,\nu(dx_1,dx_2;z/2).\cr}
        \eqno (2.32)
        $$}

        {\it Proof.} The proof goes along the same lines as
        the proof of Theorem 2.1. Moreover, in the derivation of
         the estimate
        (2.24) we used only that $T\to\infty$ and $S/T^2\to 0$
        (and not that $S/T\to\infty$), so we can apply (2.24) also
        in the present proof.
Assume that $g(x_1,x_2)\in C^\infty_0(\R^2)$.  Since
        $$
        \wL\sim S/(2R)=z/(2c)
        $$
        with $z=S/T$, $c=R/T$, (2.24) implies
        $$\eqalign{
        \liT&\IC g(P_\ep(\Rw),P_\ep(R))\,dR\cr
        &=\int_{\T^k}dt\Ic dc\, g(A_\ep(t+\om z/(2c)),A_\ep(t))\cr
        &=\Ic dc\II \II g(x_1,x_2)\,\nu_\ep(dx_1,dx_2;z/(2c)),\cr}
        $$
        where $\nu_\ep(dx_1dx_2;z)$ is a joint distribution of
        $A_\ep(t+\om z)$ and $A_\ep(t)$. Letting $\ep\to 0$, we
        come to (2.31).

        Since $\II\II g(x_1,x_2)\,\nu(dx_1,dx_2;z)$ depends
        continuously on $z$, (2.32) is a consequence of (2.31).
        Theorem 2.2 is proved.

We shall also use the following general result:

{\bf Theorem  2.3.} {\it Assume $F(R)\in B^2$. Let
$\nu(dx_1dx_2;z)$ be a probability distribution of the pair
$\quad (F(R),F(R+z))$, and $\nu(dx)$ be a probability distribution
of $F(R)$. Then for every
continuous function $g(x_1,x_2)$ which is $O(|x_1|^2
+|x_2|^2)$ at infinity,
$$
I_g(z;F)=\II\II g(x_1,x_2)\,\nu(dx_1dx_2;z)\eqno (2.33)
$$
is a continuous almost periodic function in $t$, and
$$
\liT\IO I_g(z;F) \,dz=\II\II g(x_1,x_2)\nu(dx_1)\nu(dx_2).
\eqno (2.34)
$$}

{\it Proof.} We have:
$$
I_g(z;F)=\liT\IO g(F(R),F(R+z))\,dR
$$
Assume $g(x)\in C^\infty_0(\R^2)$ and $P_\ep(x)$ is
a trigonometric polynomial satisfying
$$
\| P_\ep(R)-F(R)\|_{B^2}\le\ep.
$$
Then
$$
|I_g(z;F)-I_g(z;P_\ep )|\le C(g)\ep.\eqno (2.35)
$$
Now,
$$
P_\ep(R)=A_\ep(\om_1R,\dots,\om_kR),
$$
where $A_\ep$ is a function on $\T^k$ and $\om_1,\dots,\om_k$
are incommensurate. It implies that
$$
I_g(z;P_\ep )=\int_{\T^k}g(A_\ep(t),A_\ep(t+z\om))\,d\tau.
\eqno (2.36)
$$
Let $M=\sup_{t\in\T^k}|A_\ep(t)|.$ Consider
a polynomial $p_\ep(x_1,x_2)$ such that
$$
\sup_{|x_1|,|x_2|\le M}|p_\ep(x_1,x_2)-g(x_1,x_2)|\le\ep.
$$
Then
$$
|I_{p_\ep }(z;P_\ep ) -I_g(z;P_\ep)|\le\ep\eqno (2.37)
$$
Since $I_{p_\ep }(z;P_\ep )$ is a trigonometric
polynomial in $z$,
estimates (2.35), (2.37) prove that if $g(x)\in C_0^\infty(\R^2)$,
then $I_g(z;F)$ is an almost periodic function in $z$.

By implication it holds also for every continuous $g(x)$
which grows at infinity as $O(|x|)^2$.

 From (2.36) and the ergodic theorem,
$$
\liT\IO I_g(z;P_\ep )\,dz=\int_{\T^k}\int_{\T^k}
 g(A_\ep(t),A_\ep(s))\,dt\, ds
=\II\II g(x_1,x_2)\,\nu_\ep(dx_1)\nu_\ep(dx_2).
$$
Letting $\ep\to 0$ we obtain, with the help of (2.30),
the formula (2.34). Theorem 2.3 is proved.

        \beginsection 3. Proof of Theorems 1.1 and 1.2 \par

        {\it Proof of Theorem 1.1.}
We have:
        $$\eqalign{
        \La(R,S;\a)
        &=\La(\Rw;\a)-\La(R;\a)\cr
&=(\Rw)^{1/2}F(\Rw;\a)-R^{1/2}\FR\cr
        &=R^{1/2}(F(\Rw;\a)-\FR)
        +O(SR^{-3/2}|F(\Rw;\a)|),
        \cr}
        \eqno (3.1)
        $$
        since by (1.18),
        $$
w=\wL\sim(S/(2R)),\quad (\Rw)^{1/2}-R^{1/2}
        \sim(S/(4R^{3/2})).
        $$
Due to the triangle inequality, (3.1) implies
        $$\eqalign{
        \biggl(&\IC
        |\La(R,S;\a)|^2\dR\biggr)^{1/2}\cr
        &=\left(\IC|F(\Rw;\a)-\FR|^2R\dR\right)^{1/2}\cr
        &+O\left(\left(\IC(SR^{-3/2})^2
        |F(\Rw;\a)|^2dR\right)^{1/2}\right)\cr
        &=\left(\IC|F(\Rw;\a)-\FR|^2R\dR\right)^{1/2}
        +O(ST^{-3/2}),\cr}
        $$
        since
        $$\eqalign{
        \IC& |F(\Rw;\a)|^2dR\cr
        &={1\over(c_2-c_1)T}\int_{c_1T+w_S(c_1T)}^{c_2T+w_S(c_2T)}
        |F(R';\a)|^2(dR/dR')\,dR'=O(1),\cr}
        $$
where $R'=\Rw$. Hence
        $$\eqalign{
        &\lt0 \left(T^{-1}\,\IC|\La(R,S;\a)|^2\dR\right)^{1/2}\cr
        &=\lt0 \left(T^{-1}\,\IC|
        F(\Rw;\a)-\FR|^2R\dR\right)^{1/2},\cr}
        $$
        or equivalently,
        $$\eqalign{
        &\lt0 T^{-1}\,\IC|\La(R,S;\a)|^2\dR\cr
        &=\lt0 T^{-1}\,\IC|F(\Rw;\a)-\FR|^2R\dR.\cr}
        \eqno (3.2)
        $$
        By Theorem 2.1,
        $$\eqalign{
        &\lLT
        T^{-1}\,\IC|F(\Rw;\a)-\FR|^2R\dR\cr
        &=J_\f\lLT
        T^{-1}\,\IC|F(\Rw;\a)-\FR|^2\f_0(R/T)\,dR\cr
        &=J_\f\II\II (x_1-x_2)^2\nu_\a(dx_1)
        \nu_\a(dx_2) =J_\f W(\a),
        \cr}
        $$
where $\f_0=J_\f^{-1}c\,\f(c)$, $J_\f=\di\Ic c\,\f(c)\,dc$ and
        $$
W(\a)=2\II x^2\nu_\a(dx).
        $$
        Thus
        $$
        \lLT T^{-1}\,\IC|\La(R,S;\a)|^2dR
        =J_\f W(\a).
        $$
Now Theorem C, the Fourier expansion (1.42) and the Parceval identity (2.9)
lead to
        $$
        \II x^2\nu_a(dx)=(1/2)\pi^{-2}\sum_{k=1}^\infty|u_\a(k)|^2,
\eqno (3.3)
        $$
hence
$$
W(\a)=\pi^{-2}\Sk|u_\a(k)|^2.
$$
Theorem 1.1 is proved.

{\it Proof of Theorem 1.2.} From an analogue  of (3.2)
where $(S/T)\to z$ instead of $(S/T^2)\to 0 $ and Theorem 2.2,
$$\eqalign{
&\lLt T^{-1}\IC|\La(R,S;\a)|^2\dR\cr
&=\lLt T^{-1}\IC|F(R+\wL);\a)-F(R;\a)|^2R\dR\cr
&=\II\II |x_1-x_2|^2\Ic dc \,c \f(c)\,\nu(dx_1,dx_2;z/(2c))\cr
&=\Ic c\,\f(c)\,W(z/c;\a)\,dc\cr}
$$
with
$$
W(z;\a)=\II\II |x_1-x_2|^2\nu(dx_1,dx_2;z/2)
=\liT\IO |F(R+z/2;\a)-F(R;\a)|^2dR.
$$
Now, from (1.42),
$$\eqalign{
&F(R+z/2;\a)-F(R;\a)\cr
&=\PI\Sk |u_\a(k)|(\cos (2\pi(R+z/2)Y_k+\theta_\a(k)-\tf)-\cos(2\pi RY_k
+\theta_\a(k)-\tf))\cr
&=-2\PI\Sk u_\a(k)\sin(\pi zY_k/2)\sin(2\pi(R+z/4)Y_k+\theta_\a(k)-\tf),
\cr}
\eqno (3.4)
$$
so by the Parseval identity (2.9),
$$\eqalign{
\liT\IO|F(R+z/2;\a)-F(R;\a)|^2dR
&=2\pi^{-2}\Sk|u_\a(k)|^2\sin^2(\pi zY_k/2)\cr
&=\pi^{-2}\Sk|u_\a(k)|^2(1-\cos(\pi zY_k)),\cr
}$$
so
$$
W(z;\a)=\pi^{-2}\Sk |u_\a(k)|^2(1-\cos(\pi zY_k)).
$$
Theorem 1.2 is proved.

        \beginsection 4. Proof of Theorems 1.3, 1.4 \par

        {\it Proof of Theorem 1.3.} Assume first that $\g\in\G_0$.
Then $Y(m)\not= Y(n)$ for $m\not= n$, hence (1.33) reduces to
        $$\eqalign{
W(z;\a)&=\pi^{-2}\sz |n|^{-3}\rho(n) (1-\cos(\pi Y(n)z))\cr
&=2\pi^{-2}\sz|n|^{-3}\rho(n)\sin^2(\pi Y(n)z/2),\cr}
\eqno (4.1)
$$
so that
$$
z^{-1}W(z;\a)
        =\pi^{-1}
        \sz|\pi nz/2|^{-3}\rho(\pi nz/2)
        \sin^2(Y(\pi nz/2))\,(\pi z/2)^2.
        $$
        Here we used the fact that $\rho(\la\xi)=\rho(\xi)$
        and $Y(\la\xi)=\la Y(\xi)$ for every $\la>0$, and
        we reduced $z^{-1}W(z;\a)$ to an approximating sum to the integral
        $$
\pi^{-1}\II\II|\xi|^{-3}\rho(\xi)
        \sin^2Y(\xi)\,d\xi.
        $$
        In the Appendix  we shall show that
        $$
        \pi^{-1}\II\II|\xi|^{-3}\rho(\xi)
        \sin^2Y(\xi)\,d\xi=1, \eqno (4.2)
$$
        so
        $$
        \lim_{z\to 0}z^{-1}W(z;\a)=1,
        $$
        which was stated.

Assume now that $\g\in\G_0(H)$. For the sake of simplicity we will consider
$H=\{\text{Id},g_1\}$, where $g_1\: (x_1,x_2)\to (x_1,-x_2)$. The
general case is treated in the same way.
In the case under consideration
$$
m_\a(H)=\quad
\eqalign{
&2 \quad \text{if}\quad \a_2 \;\;\text{is half--integer},\cr
&1 \quad \text{if}\quad \a_2 \;\;\text{is not half--integer},\cr
}$$
so we have to show that
$$
\lim_{z\to 0}z^{-1}W(z;\a)=\quad
\eqalign{
&2 \quad \text{if}\quad \a_2 \;\;\text{is half--integer},\cr
&1 \quad \text{if}\quad \a_2 \;\;\text{is not half--integer}.\cr
}$$
Define
$$
P=\{(n_1,n_2)\in\Z^2\: n_2=0\},
$$
so that $P$ consists of fixed points of $g_1$. From (1.26),
if $Y(n)=Y_k$
then
$$
|u_\a(k)|=\quad
\eqalign{
&2|\cos(n_2\a_2)|\,\NS,\quad\text{if}\quad n\not\in P,\cr
&\NS,\quad\text{if}\quad n\in P,\cr}
$$
so (1.33) reduces to
$$\eqalign{
W(z;\a)&=
4\pi^{-2}\sum_{n\not\in P}\cos^2(2\pi n_2\a_2)\NSS\sin^2(\pi Y(n)z/2)\cr
&+2\pi^{-2}\sum_{n\in P\setminus\{0\}}\NSS\sin^2(\pi Y(n)z/2)\cr
&=4\pi^{-2}\sz\cos^2(2\pi n_2\a_2)\NSS\sin^2(\pi Y(n)z)\cr
&-2\pi^{-2}\sum_{n\in P\setminus\{0\}}\NSS\sin^2(\pi Y(n)z/2)\cr
&=W_0(z;\a)-W_1(z;\a)\cr}
\eqno (4.3)
$$
with
$$\eqalignno{
&W_0(z;\a)=4\pi^{-2}\sz\cos^2(2\pi n_2\a_2)\NSS\sin^2(\pi Y(n)z/2),&(4.4)\cr
&W_1(z;\a)=2\pi^{-2}\sum_{n_1\in \Z\setminus\{0\}}\NSS\sin^2(\pi Y(n)z/2),\quad
   n=(n_1,0).
&(4.5)
\cr}
$$
Assume first that $\a_2=0$ or 1/2. Then
$$
W_0(z;\a)=4\pi^{-2}\sz\NSS\sin^2(\pi Y(n)z/2),
$$
which is twice bigger than the sum in (4.1), so the same computation
as before leads to
$$
\lim_{z\to 0}z^{-1}W_0(z;\a)=2. \eqno (4.6)
$$
Now,
$$
\sum_{n_1=1}^{z^{-1}}n_1^{-3}\sin^2(\pi Y((n_1,0))z)
\le Cz^2\sum_{n_1=1}^{z^{-1}}n_1^{-1}\le C_0z^2|\log z|,
$$
and
$$
\sum_{n_1=z^{-1}}^{\infty}n_1^{-3}\sin^2(\pi Y((n_1,0))z)
\le \sum_{z^{-1}}^\infty n_1^{-3}\le C_1 z^2,
$$
hence
$$
0\le W_1(z;\a)\le C z^2|\log z|.\eqno (4.7)
$$
 From (4.3)--(4.7),
$$
\lim_{z\to 0}z^{-1}W(z;\a)=2,
$$
which was stated.

Consider now the case when $\a_2$ is not half--integer.
        Substituting
$$
\cos^2(2\pi n_2\a_2)=(1+\cos(4\pi n_2\a_2))/2
$$
        into (4.4), we obtain that
        $$
        W_0(z;\a)=W_2(z;\a)+W_3(z;\a)
        $$
        with
        $$
W_2(z;\a)=2\pi^{-2}\sz|n|^{-3}\rho(n) \sin^2(\pi Y(n)t/2)
        $$
        and
        $$
W_3(z;\a)=2\pi^{-2}\sz\cos(4\pi n_2\a_2)|n|^{-3}\rho(n)
        \sin^2(\pi Y(n)z/2).
        \eqno (4.8)
        $$
        Since $W_2(z;\a)$ coincides with $W(z;\a)$ in (4.1),
        we obtain that
        $$
        \lim_{z\to 0}z^{-1}W_2(z;\a)=1.
        $$
        Let us prove that
        $$
        \lim_{z\to 0}z^{-1}W_3(z;\a)=0.
        \eqno (4.9)
        $$
The idea is to apply the Abel summation
        formula to (4.9). Let us fix $n_1=j$ and denote
        $$
        a(k)=\cos (4\pi k\a_2)
        $$
        and
        $$
        b_j(k)=|n|^{-3}\rho(n)\sin^2(\pi Y(n)z/2), \quad n=(j,k).
        $$
We have:
        $$
        \sum_{k=-\infty}^\infty a(k)b_j(k)
        =\sum_{k=-\infty}^\infty A(k)(b_j(k)-b_j(k+1))
        $$
        with
        $$
        A(k)=\quad \quad\eqalign{
        &a(1)+\dots+a(k),\quad k\ge 1,\cr
        &0,\quad k=0,\cr
        &-a(k+1)-\dots -a(0),\quad k\le -1.}
        $$
        Since $\a_2$ is not half--integer,
        $$
        |A(k)|=
        \left|\sum_{i=i_0(k)}^{i_1(k)}\cos (4\pi i\a_2)\right|
        $$
        is uniformly bounded in $k$. In addition,
        $$
        |b_j(k)-b_j(k+1)|\le C(|n|^{-4}\sin^2(\pi Y(n)z/2)
        +|n|^{-3}z|\sin(\pi Y(n)z/2)|),
        $$
        so
        $$
        |W_3(z;\a)|\le C_0
        \sz (|n|^{-4}\sin^2(\pi Y(n)z/2)
        +|n|^{-3}z|\sin (\pi Y(n)z/2)|).
\eqno (4.10)
        $$
        Let us show that
        $$\eqalign{
        &\sz |n|^{-4}\sin^2(\pi Y(n)z/2)\le Cz^2|\log z|,\cr
        &\sz |n|^{-3}z|\sin (\pi Y(n)z/2)|\le Cz^2|\log z|.\cr}
        \eqno (4.11)
        $$
        We have:
        $$
        \sum_{0\le|n|\le 1/z}
         |n|^{-4}\sin^2(\pi Y(n)z/2)
        \le \sum_{0\le|n|\le 1/z}
        C|n|^{-2}z^2\le C_1 z^2|\log z|
        $$
        and
        $$
        \sum_{|n|\ge 1/z}
         |n|^{-4}\sin^2(\pi Y(n)z/2)
        \le \sum_{|n|\ge 1/z}
        |n|^{-4}\le C_1 z^2
        $$
        which proves the first part of (4.11). The second part
        is established in the same way.

 From (4.10), (4.11),
$$
W_3(z;\a)=O(z^2|\log z|),\quad z\to 0,
$$
which proves (4.9).
 From (4.9),
$$
\lim_{z\to 0} z^{-1}W(z;\a)=1,
$$
which was stated. Theorem 1.3 is proved.

{\it Proof of Theorem 1.4.} When $\g$ is the circle
$\{|x|=\pi^{-1/2}\}$ of area 1, then
$Y(n)=\pi^{-1/2}|n|$, $\rho(n)=\pi^{-1/2}$ and
(1.33) reduces to
$$
W(z;\a)=2\pi^{-5/2}
\sum_{k=1}^\infty
|r_\a(k)|^2k^{-3/2}\sin^2(\pi^{1/2} k^{1/2}z/2)
\eqno (4.12)
$$
with
$$
r_\a(k)=\sum_{n\in\Z^2\: n^2_1+n_2^2=k}e(n\a).
$$
Define
$$
\psi(\xi)=\xi^{-3/2}\sin^2(\xi^{1/2}),\quad \De=\pi z^2/4.
\eqno (4.13)
$$
Then (4.12) is equivalent to
$$
z^{-1}W(z;\a)=\pi^{-2}\Sk|r_\a(k)|^2\psi(k\De)\De.
\eqno (4.14)
$$
We shall use the following result from [BD1]:

{\bf Theorem D.} {\it For all rational $\a\in\Q^2$,
$$
\lim_{N\to\infty}(N\log N)^{-1}
\sum_{k=1}^N|r_\a(k)|^2=C(\a)>0.
\eqno (4.15)
$$
For all Diophantine $\a$,
$$
\lim_{N\to\infty}N^{-1}
\sum_{k=1}^N|r_\a(k)|^2=\pi.
\eqno (4.16)
$$}

{\it Remark.} Actually in [BD1] an explicit formula
was derived for $C(\a)$. For $\a=0$ it gives
$C(0)=3$.

Consider
$$
I(a,a+\ep;\De)
=\sum_{k\: a\le k\De <a+\ep}
|r_\a(k)|^2\psi(k\De)\De,\quad 0<\De \ll \ep\ll a.
$$
Let us  replace $\psi(k\De)$ by $\psi(a)$ in the RHS
of the last formula and estimate
the error coming from this replacement:
$$
|I(a,a+\ep;\De)-\psi(a)S(a,a+\ep;\De)|
\le C\ep\psi_0(a)S(a,a+\ep;\De),
\eqno (4.17)
$$
where
$$
S(a,a+\ep;\De)=
\sum_{k\: a\le k\De <a+\ep}|r_\a(k)|^2\De,
$$
and
$$
\psi_0(\xi)=\xi^{-3/2}.
$$
Assume $\a\in\Q^2$. Then from (4.15),
$$\eqalign{
S(a,a+\ep;\De)
&=C(\a)
((a+\ep)\log((a+\ep)/\De)-
a\log(a/\De))+o(|\log\De|)\cr
&=C(\a)\ep|\log\De|+o(|\log\De|),\quad \De\to 0.\cr}
\eqno (4.18)
$$
(4.17) and (4.18) imply
$$
|I(a,a+\ep;\De)-C(\a)|\log\De|\psi(a)\ep|
\le C|\log\De|\psi_0(a)\ep^2,\quad \De\le\De_0(\ep).
\eqno (4.19)
$$
Summing up this estimate for $N$ adjacent intervals
$[a+j\ep,a+(j+1)\ep]$, $j=0,\dots,N-1$,
we obtain that
$$
\left|I(a,b;\De)-
C(\a)|\log\De|\sum_{j=0}^{N-1}\psi(a+j\ep)\ep\right|
\le C|\log\De|\sum_{j=0}^{N-1}\psi_0(a+j\ep)\ep^2,
\quad b=a+N\ep,
$$
which implies
$$
\left|I(a,b;\De)-C(\a)|\log\De|\int_a^b \psi(t)dt\right|
\le C_0|\log\De|\ep, \quad C_0=C_0(a).\eqno (4.20)
$$
It remains to estimate $I(0,a;\De)$ and $I(b,\infty;\De)$.

Assume $a<1$. Then $\psi(\xi)$  is a decreasing function
on $(0,a)$,
hence
$$
I(a/2,a;\De)\le\psi(a/2)\,S(a/2,a;\De).
$$
Now,
$$
S(a/2,a;\De)\le S(0,a;\De)
=\sum_{k\: k\De<a}|r_\a(k)|^2\De
\le Ca\log(a/\De)\le Ca|\log\De|,
$$
so
$$
I(a/2,a;\De)\le C_0|\log\De|\psi(a/2)\,(a/2).
$$
Replacing $a$ by $a2^{-j}$, we obtain that
$$
I(a2^{-j-1},a2^{-j};\De)\le C_0|\log\De|\psi(a2^{-j-1})\,a2^{-j-1},
\quad j=0,1,\dots, J,
\eqno (4.21)
$$
where
$$
J=\min\{ j\:a2^{-j}\le \De\}.
$$
Summing up (4.21) in $j=0,1,\dots,J$, we obtain that
$$
I(0,a;\De)\le C_1|\log \De|\int_0^a\psi(\xi)\,d\xi.
\eqno (4.22)
$$
Assume $b>1$. Then along the same way we obtain that
$$
I(b,2b;\De)\le C_0|\log\De|\psi_1(b)b,
\quad \psi_1(\xi)=\xi^{-1/2}(1+\xi^{-1}),
$$
and then that
$$
I(b,\infty;\De)\le C_1|\log\De|\int_b^\infty\psi_1(\xi)\,d\xi.
\eqno (4.23)
$$
Choosing $a\ll 1$ and $b\gg 1$, and then $\ep\ll\min\{a,C^{-1}_0(a)\}$,
where $C_0(a)$ is the constant in (4.20), we obtain from
(4.20), (4.22) and (4.23) that
$$
I(0,\infty;\De)=C(\a)|\log\De|\int_0^\infty\psi(\xi)\,d\xi
+o(|\log\De|),\quad \De\to 0.
\eqno (4.24)
$$
By (4.14), $z^{-1}W(z;\a)=\pi^{-2}I(0,\infty;\De)$. In addition,
$$
\int_0^\infty\psi(\xi)\,d\xi=\int_0^\infty\xi^{-3/2}\sin^2\xi^{1/2}d\xi
=2\int_0^\infty\eta^{-2}\sin^2\eta\, d\eta=\pi,
$$
therefore
$$
z^{-1}W(z;\a)=C(\a)\pi^{-1}|\log\De|+o(|\log\De|),
$$
or, since $\De=\pi z^2/4$,
$$
z^{-1}W(z;\a)=2C(\a)\pi^{-1}|\log z|+o(|\log z|),
$$
hence
$$
\lim_{z\to 0}(z|\log z|)^{-1}W(z;\a)=2C(\a)\pi^{-1}.
$$
Thus the first part of Theorem 1.4 is proved.

Assume now that $\a$ is Diophantine. Then by (4.16),
$$
S(a,a+\ep;\De)=\pi\ep+o(1),\quad \De\to 0,
$$
hence (4.17) implies that
$$
|I(a,a+\ep;\De)-\pi\ep|\le C\psi_0(a)\ep^2,\quad
\De\le\De_0(\ep).
$$
Now, along the same way as we derived (4.24) for
rational $\a$, we arrive at
$$
I(0,\infty;\De)=\pi\int_0^\infty\psi(\xi)\,d\xi+o(1)=\pi^2+o(1),\quad\De\to 0.
$$
This implies
$$
z^{-1}W(z;\a)=\pi^{-2}I(0,\infty;\De)\,=\,1+o(1),\quad z\to 0,
$$
which proves Theorem 1.4 for Diophantine $\a$.

\beginsection 5. Proof of Theorems 1.5--1.7 \par

{\it Proof of Theorem 1.5.} By (3.1)
$$
\FRL=F(R+\wL;\a)-\FR+O(SR^{-2}|F(R+\wL;\a)|),
$$
hence assuming $g(x)\in C_0^\infty(\R^1)$, we obtain that
$$\eqalign{
&\IC g(\FRL)\dR\cr
&=\IC g(\FW-\FR)\dR+O(ST^{-2}).\cr} \eqno (5.1)
$$
By Theorem 2.1,
$$\eqalign{
&\lLT\IC g(\FW-\FR)\dR\cr
&=\II\II g(x_1-x_2)\nu_\a(dx_1)\nu_\a(dx_2)=\II g(x)\mu_\a(dx),\cr}
$$
hence
$$
\lLT\IC g(\FRL)\dR=\II g(x)\mu_\a(dx).
$$
By implication this equation holds for every continuous
bounded function $g(x)$.
Theorem 1.5 is proved.

{\it Proof of Theorem 1.6.} From (5.1) and Theorem 2.2
we obtain that
$$\eqalign{
&\lLt\IC g(\FRL)\dR\cr
&=\II\II g(x_1-x_2)\Ic dc\,\f(c)\,
\nu_\a(dx_1dx_2;z/(2c)),\cr}
$$
where $\nu_\a(dx_1,dx_2;z)$ is a joint distribution of
$\FR$ and $F(R+z;\a)$. Let $\mu_\a(dx;z)$ be a probability
distribution of a difference $\xi_1-\xi_2$ of two random
variables, whose joint distribution coincides with
$\nu_\a(dx_1dx_2;z/2)$. Then
$$
\II\II g(x_1-x_2)\nu_\a(dx_1dx_2;z/2)
=\II g(x)\mu_\a(dx;z),
\eqno (5.2)
$$
hence
$$
\lLt\IC g(\FRL)\dR
=\II g(x)\mu_\a(dx;z).
$$
Theorem 1.6 is proved.

{\it Proof of Theorem 1.7.} Almost periodicity
of
$$
\II g(x)\,\mu_\a(dx;z)
$$
follows from the equation (5.2) and the first part
of Theorem 2.3. The value of
$$
\lim_{T\to\infty}\IO dz\II g(x)\,\mu_\a(dx;z)
$$
follows from (5.2) and the second part of Theorem 2.3.
Theorem 1.7 is proved.

\beginsection 6. Proof of Theorems 1.8, 1.9 \par

{\it Proof of Theorem 1.8.} For the sake of simplicity we will assume
   $\g\in\G_1$. The case
   $\g\in\G_1(H)$ is treated similarly.
By (5.2) $\mat$ is a distribution of
the almost periodic function $F(R+z/2;\a)-F(R;\a)$. From (1.41),
$$\eqalign{
&F(R+z/2;\a)-\FR\cr
&=-2\PI\sz e(n\a)\NS\sin(\pi zY(n)/2)
\sin(2\pi(R+z/4)Y(n)-3\pi/4).\cr}
\eqno (6.1)
$$
Define for every $n\in M$,
$$
f_n(s;z,\a)
=-2\PI\Sk e(kn\a)|kn|^{-3/2}\sqrt {\rho(n)}
\sin(\pi zY(kn)/2)
\sin(2\pi ks+\pi zY(kn)/2-3\pi/4),
\eqno (6.2)
$$
which is a periodic function in $s$ of period 1. Then (6.1) can
be rewritten as
$$
F(R+z/2;\a)-\FR=
\sum_{n\in M} f_n(Y(n)R;z,\a).
$$
By our assumption, the numbers $Y(n),\;n\in M$ are linearly
independent over $\Q$, so by Lemmas 4.4 and 2.5  in [Bl3]
$\matt$,
the distribution of $F(R+z/2;\a)-\FR$, coincides with the distribution
of the random series
$$
\xi_{z\a}=\sum_{n\in M} f_n(t_n;z,\a),\eqno (6.3)
$$
where $t_n$ are independent random variables, uniformly
distributed on $[0,1]$.
Since $\g\in \G_1\subset \G_0$, all the numbers
$\{ \, Y(n),n\in M \, \}$ are different. Let us order the
numbers $\{ \,Y(n),n\in M \, \} $ in the increasing order, i.e.,
$ Y(n(1))\,<\, Y(n(2))< \dots $
Denote $$a_k(s,z,\alpha )=f_{n(k)}(s;z,\a ), \quad
a_k(s)=a_k(s;z,\a).\eqno (6.4)$$
According to Theorem 3.1 in [Bl3], Theorem 1.8
will follow,
except the lower bound (1.39), if we prove that
$$
\eqalign {\sup _{0\le s \le 1} |a_k(s)| \, &< \,
Jk^{-3/4},\quad J>0;\cr
\sum_{j=k}^\infty  \int _0^1 a_j(s)^2 ds \,
&>\, J_0k^{-1/2},\quad J_0>0. \cr}\eqno (6.5)
$$
Since $Y(\xi )$ is a positive homogeneous function of second order,
$$C_0k^{1/2}\le |n(k)|\le C_1 k^{1/2}, \quad C_0,C_1 >0.
\eqno (6.5')$$
Hence (6.5) is equivalent to the following two estimates:
$$
\sup _{0\le s\le 1} |f_n(s;z,\a)|\,<\, J_1 |n|^{-3/2},\quad J_1>0, \eqno (6.6)
$$
$$
\sum_{n\in M\: |n|>r} \int _0^1 (f_n(s;z,\a))^2 ds \, >J_2 r^{-1}, \quad J_2>0.
   \eqno (6.7)
$$
The first estimate follows from (6.2):
$$
|f_n(s;z,\a)|\le C\sum _{k=1}^\infty |kn|^{-3/2}\le C_0 |n|^{-3/2}.
$$
Let us prove (6.7). (6.2) is a Fourier series in $s$, so
$$
\int_0^1 (f_n(s;z,\a))^2ds=2\pi^{-2}\sum_{k=1}^\infty |kn|^{-3} \r
(n)\sin^2(\pi
    z Y(kn)/2)
\ge C|n|^{-3} \sin ^2(\pi z Y(n)/2). \eqno(6.8)
$$
and
$$
\sum _{n\in M:|n|>r}\int_0^1 (f_n(s;z,\a))^2ds
\ge C\sum_{n\in M\: |n|>r} |n|^{-3} \sin^2(\pi z Y(n)/2). \eqno (6.9)
$$
Note that the density of $M$ in the plane is $6/\pi ^2 $, i.e.,
$$
\lim _{r\to\infty}(\pi r^2)^{-1}\sum_{n\in M\: |n|<r} 1 =6/\pi ^2,
$$
so (6.7) will follow from (6.9) if we show that $\forall \ep >0 \quad \exists
\d
   e >0$ such that
the upper density of the set
$Q_\de=\{n\in \Z^2\, :\, \sin^2(\pi z Y(n)/2) \le \de ^2\,\}
$
is less than $\ep $. Indeed, then for large $r$,
$$
\sum_{n\in M:|n|>r}
\int _0^1 (f_n(s;z,\a))^2 ds
\ge C \sum_{n\in M\setminus Q_\de  :|n|>r}|n|^{-3} \de ^2
\ge C((6/\pi ^2)-2\ep)\de ^2
\sum_{n\in \Z ^2 :|n|>r}|n|^{-3} \ge C_0r^{-1},
$$
which gives (6.7).
Let $\sin \de _0 =\de $. Consider the annuli
$$
A_\de (l) =\{x\in \R^2\,: \, |\pi zY(x)/2 -l|<\de_0\,\}
$$
and define
$
Q_\de (l) = \Z^2 \cap A_\de (l).
$
Then
$
\di Q_\de =\di \cup_{l=1}^\infty Q_\de (l).
$
Let $ N_\de (l) =|Q_\de (l)|.  $
Due to the $2/3$--estimate of Sierpinski (see, e.g., [R] or [CdV]),
$$
|N_\de (l)-\text { Area } A_\de (l)|\le Cl^{2/3}
$$
In addition, $\text { Area } A_\de (l) \le C_0\de l $, hence
$N_\de (l)\le C_0\de l +Cl^{2/3}$ and
$$
\sum _{l=1}^L N_\de (l) \le C_0\de L^2 +CL^{5/3}
$$
Therefore the upper  density $d (Q_\de )$ of $Q_\de $ is estimated as
$$
d(Q_\de )\ \le \ \ls _{L\to\infty} C_1L^{-2} \sum_{l=1}^L N_\de (l) \le C_2 \de
   .
\eqno (6.10)
$$
This proves the desired estimate of  $d(Q_\de ) $, and so Theorem 1.8 is
proved,
except (1.39). To prove (1.39) we use the following theorem.

{\bf Theorem 6.1.} {\it Assume $a_k(s),\ k=1,2,\dots,$ are continuous functions
   of period $1$,
with $\di \int_0^1 a_k(s)ds =0$ and
$$
\ls _{k\to\infty} \sup_{0\le s\le 1} |a_k(s)|\, <\, \infty, \quad
\sum_{k=1}^\infty \int_0^1 |a_k(s)|^2 ds <\infty.
$$
Assume also that  $\exists \;k_0$ and $\de_0>0$ such
that   $\forall k>k_0\ \exists\ G_k\subset \{1,2,\dots,k\}$
such that $|G_k|>\de_0k$
and $\forall l\in G_k$,
$$C' l^{-\g}>\sup _{0\le s \le 1} |a_l(s)|\ge
\left (\int_0^1 |a_l(s)|^2ds\right )^{1/2}
\ge C''l^{-\g},\eqno(6.11)$$
with some $C',C''>0$ and $0<\g < 1.$

Then $\forall x\ge 0$,
$$
\eqalignno{
&\Pr \, \left \{\, \sum_{k=1}^\infty a_k(t_k)\,>\,x\,\ \right \}\,>\,
C\exp(-\la x^{1/(1-\g ) })&(6.12)\cr
&\Pr \, \left \{\, \sum_{k=1}^\infty a_k(t_k)\,<\,-x\,\right \}\, >\,
C\exp(-\la x^{1/(1-\g )})&(6.13)\cr }
$$
with some $C, \la>0$,
assuming that $t_k\in [0,1],k\ge 1,$ are independent
uniformly distributed random variables.

Proof. } The proof follows the proof of Theorem 5.1 in [BCDL] and Theorem 3.3
in
    [Bl3]. Define
$$
A_l=\{s\, :a_l(s)\ge \de_1l^{-\g }\,\},\quad\de_1>0.
$$
Then (6.11) and $\di \int_0^1a_l(t)dt=0$ imply that
$\exists\de_1,\de_2>0:$
$\forall l\in G_k $, $\text { mes } A_l>\de_2$
(otherwise $L^2$--norm of $a_l(t)$ is too small). For $l\not \in G_k$ define
$$
A_l=\{t\,:\, a_l(t)\ge -l^{-5}\,\}
$$
Again $\mes A_l >l^{-10}$, $l\ge l_0 $,  (otherwise $\di \int_0^1 a_l(t)dt
<0$).
Define $$D_k=\{(t_1,t_2,\dots )\,:\, t_l\in A_l,\, l=k_0,\dots,k\,;
\sum_{l=k+1}^\infty |a_l(t_l)| <1 \, \}. $$
For large $k$,
$$
\Var \sum_{l=k+1}^\infty a_l(t_l))=
\sum_{l=k+1}^\infty \int_0^1 |a_l(s)|^2 ds <1/2,
$$
so by Chebyshev's inequality
$$
\Pr \, \left \{\, \sum_{l=k+1}^\infty a_l(t_l)\,>\, 1 \, \right \}\,\le 1/2,
$$
so
$$\Pr D_k \ge (1/2) (\de_2)^k  \ge \exp (-\la_0k),\quad\la_0>0.
\quad k\ge k_1(\ep ).$$
For $(t_1,t_2,\dots )\in D_k$,
$$\sum_{l=1}^\infty a_l(t_l)
\ge \left (\de_1\sum_{l\in G_k} l^{-\g-4\ep } \right )
-C \ge \de_0\de_1k^{1-\g}-C=\ge\de_3k^{1-\g},
\quad k\ge k_2.$$
Thus
$$\Pr \, \left \{\, \sum_{l=1}^\infty a_l(t_l)\,\ge \,x\,=
\de_3k^{1-\g}\ \right \}\ge
 \Pr D_k \ge  \exp (-\la_0k)=
\exp(-\la x^{1/(1-\g)}), \ x\ge x_0,
$$
which proves (6.12).
(6.13) is established along the same way. Theorem 6.1 is proved.

{\bf Lemma 6.2. } {\it $\exists \de \, >\, 0$ such that
 $a_k(s)=f_{n(k)}(s;z,\a)$ satisfies the condition of Theorem 6.1 with
$G_k=\{\, 1\le l \le k \: n(l)\in M\setminus Q_\de \, \} $ and $\g=3/4.$ }

{\it Proof.} The first estimate in (6.11) follows from (6.5).
To prove the second estimate, remark that by (6.8) and ($6.5'$),
$$
\int_0^1(f_{n(l)}(s;z,\a))^2ds\ge Cl^{-3/2}\sin^2
(\pi zY(n(l))/2).
$$
If $n(l)\in M\setminus Q_\de$, then $\sin^2(\pi zY(n(k))/2)\ge\de^2$,
so
$$
\int_0^1(f_{n(l)}(s;z,\a))^2ds\ge Cl^{-3/2}\de^2,
$$
which proves the second estimate in (6.11). From (6.10)
we obtain that
$$
|G_k|\ge \de_0 k,\quad \de_0>0.
$$
Lemma 6.2 is proved.

Theorem 6.1 and Lemma 6.2 imply (1.39),
so  Theorem  1.8 is proved. Proof of Theorem 1.9 goes along
the same lines, so we omit it.

\beginsection 7. Proof of Theorems 1.10, 1.11 \par

{\it Proof of Theorem 1.10.} Assume  $\g\in\G_1$. As was shown in
Section 6,
 $\mat$ is a distribution of
the  random series (6.3) so to prove
Theorem 1.10 we have to prove that the distribution of
$\xi_{z\a}/\sqrt{\Var\xi_{z\a}}$ converges to a standard
normal distribution as $z\to 0$. To that end we shall
check that the Lindeberg condition holds for the random
series (6.3).

By the Parseval formula,
$$\eqalign{\Var \xi_{z\a}=
\II x^2\mat
&=\|F(R+z/2;\a)-\FR\|_{B^2}^2\cr
&=2\pi^{-2}\sz |n|^{-3}\rho(n)\sin^2(\pi zY(n)/2).\cr}
$$
By (4.1) this coincides with $W(z;\a)$, so by Theorem 1.3,
$$
\Var \xi_{z\a }
=z+o(z),\quad z\to 0.\eqno (7.1)
$$
Let
$$
\sigma_{z\a}(n)=(\Var f_n(t_n;z,\a))^{1/2}
=\left(\int_0^1 |f_n(s;z,\a)|^2ds\right)^{1/2}.
$$
Then the Lindeberg condition is that for every $n\in M$,
$$
\lim_{z\to 0}\sigma_{z\a}^2(n)/\Var\xi_{z\a}=0,
\eqno (7.2)
$$
and for every $\ep>0$,
$$
g_{z\a}(\ep)=(\Var\xi_{z\a})^{-1}\sum_{n\in M}
\int_{|s|\ge\ep (\Var \xi_{z\a})^{1/2}}s^2f_n(s;z,\a)ds \to 0
\eqno (7.3)
$$
as $z\to 0$. From (6.2),
$$
|\fs|\le C|n|^{-3/2}\Sk k^{-3/2}|\sin(\pi zY(kn)/2)|.
$$
It implies that
$$
|\fs|\le\quad\eqalign{
&C_0|n|^{-1}z^{1/2}\quad\text{when}\quad |n|z<1,\cr
&C_0 |n|^{-3/2}\quad\text{when}\quad |n|z>1.\cr}
\eqno (7.4)
$$
Indeed, if $|n|z<1$ then
$$
\Sk k^{-3/2}|\sin(\pi zY(kn)/2)|
=\De^{1/2}\Sk |k\De|^{-3/2}\sin(k\De)\De\le C_1 |n|^{1/2}z^{1/2},
\quad \De=\pi zY(n)/2,
$$
which proves the first part of (7.4). The second part follows from
the evident inequality
$$
\Sk k^{-3/2}|\sin(\pi zY(kn)/2)|\le\Sk k^{-3/2}.
$$
Similarly,
$$
\sigma_{z\a}^2(n)=\int_0^1|\fs|^2ds\le C|n|^{-3}\Sk k^{-3}
|\sin (\pi zY(kn)/2)|^2,
$$
hence
$$
\sigma_{z\a}^2(n)\le\quad\eqalign{
&C_0|n|^{-1}z^{2}|\log z|\quad\text{when}\quad |n|z<1,\cr
&C_0 |n|^{-3}\quad\text{when}\quad |n|z>1.\cr}
\eqno (7.5)
$$
Comparing (7.1) with (7.5), we obtain (7.2).

To prove (7.3) remark, that (7.4) implies that
for every $\ep>0$ the inequality
$$
\sup_{0\le s\le 1}|\fs|\ge\ep(\Var\xi_{z\a})^{1/2}
=\ep z^{1/2}(1+o(1))
$$
holds only for a finite set $n\in M_\ep\subset M$,
which does not depend on $z$. By (7.3),
$$
g_{z\a}(\ep)\le (\Var\xi_{z\a})^{-1}\sum_{n\in M_\ep}\sigma_{z\a}^2(n),
$$
hence the condition $ \lim_{z\to 0}g_{z\a}(\ep)=0 $
follows from (7.2).

Thus the random series (6.3) satisfies the Lindeberg
condition, which implies that the distribution of
$\xi_{z\a}/\sqrt{\Var
\xi_{z\a}}$ converges to a standard normal distribution
as $z\to 0$ (see, e.g., [L]). This proves Theorem 1.10 in the case when
$\g\in \G_1$. The same arguments work in the case
when $\g\in\G_1(H)$, so Theorem 1.10 is proved.

{\it Proof of Theorem 1.11.}
For $\g=\{|x|=\pi^{-1/2}\}$, (1.41) reduces to
$$\eqalign{
\FR
&=\PI\sz|n|^{-3/2}\pi^{-1/4}\cos(2\pi^{1/2}R|n|+\phi(n;\a))\cr
&=\Re\left\{\pi^{-5/4}\sz|n|^{-3/2}\exp(2\pi^{1/2}R|n|i+\phi(n;\a)i)\right\}\cr
&=\pi^{-5/4}\Sk r_\a(k)k^{-3/4}\cos(2\pi^{1/2}Rk^{1/2}-3\pi/4)
\cr}
\eqno (7.6)
$$
with
$$
r_\a(k)=\sum_{n\in\Z^2\: n_1^2+n_2^2=k}e(n\a).
\eqno (7.7)
$$
hence
$$\eqalign{
F(R+z/2;\a)-&\FR\cr
&=-2\pi^{-5/4}\Sk r_\a(k)k^{-3/4}\sin(\pi^{1/2}zk^{1/2}/2)
\sin(2\pi^{1/2}(R+z/4)k^{1/2}-3\pi/4)\cr
&=\sum_{\text{square free}\; k}f_k(k^{1/2}R;z,\a)\cr}
\eqno (7.8)
$$
with
$$
f_k(s;z,\a)=-2\pi^{-5/4}\Sl r_\a(l^2k)\,(l^2k)^{-3/4}
\sin(\pi^{1/2}z(l^2k)^{1/2}/2)
\sin(2\pi^{1/2}(ls+zlk^{1/2})/4-3\pi/4).
\eqno (7.9)
$$
$k\in\N$ is square free if $k=k'l^2$ implies $l=1$.
The numbers \{$k^{1/2}$, $k$ is square free\} are linearly independent over
$\Q$
   , so $\mat$,
the distribution of $F(R+z/2;\a)-\FR$, coincides with the distribution
of the random series
$$
\xi_{z\a}=
\sum_{\text{square free}\; k}f_k(k^{1/2}t_k;z,\a),\eqno (7.10)
$$
where $t_k$ are independent random variables, uniformly
distributed on $[0,1]$ (see [BCDL] and [Bl3]).
Let us check the Lindeberg condition for the random
series (7.10) as $z\to 0$.

 From (7.8),
$$\eqalign{
\Var \xi_{z\a}
&=\II x^2\mat=\|F(R+z/2)-\FR\|_{B^2}^2\cr
&=2\pi^{-5/2}\Sk|r_\a(k)|^2k^{-3/2}\sin^2(\pi^{1/2}zk^{1/2}/2)\cr
&\ge C_0\sum_{k=1}^{z^{-2}}|r_\a(k)|^2k^{-1/2}z^2
\ge C_0 z^3\sum_{k=1}^{z^{-2}}|r_\a(k)|^2.\cr}
\eqno (7.11)
$$
By Theorem D in [BCDL],
$$
\sum_{k=1}^{z^{-2}}|r_\a(k)|^2\ge z^{-2}|\log z|^{-1},\quad z\le z_0(\a),
\eqno (7.12)
$$
hence
$$
\Var\xi_{z\a}\ge Cz|\log z|^{-1},\quad z\le z_0(\a).
\eqno (7.13)
$$

Let
$$
\sigma_{z\a}(k)=(\Var f_k(t_k;z,\a))^{1/2}
=\left(\int_0^1 |f_k(s;z,\a)|^2ds\right)^{1/2}.
$$
Then the Lindeberg condition is that for every $k\in \N$,
$$
\lim_{z\to 0}\sigma_{z\a}^2(k)/\Var\xi_{z\a}=0,
\eqno (7.14)
$$
and for every $\ep>0$,
$$
g_{z\a}(\ep)=(\Var\xi_{z\a})^{-1}\Sk
\int_{|s|\ge\ep (\Var \xi_{z\a})^{1/2}}s^2f_k(s;z,\a )ds \to 0
\eqno (7.15)
$$
as $z\to 0$. Since
$$
|r_\a(k)|\le C_\de k^\de,\quad\forall\,\de>0,
$$
(see, e.g., [HW]) we obtain from (7.9) that
$$
|\fk|\le
C_\de \Sl(l^2k)^{-(3/4)+\de}|\sin(\pi^{1/2} zlk^{1/2}/2)|.
$$
It implies that
$$
|\fk|\le\quad\eqalign{
&C_0(\de) z^{(1/2)-2\de}k^{-1/2}\quad\text{when}\quad k^{1/2}z<1,\cr
&C_0(\de) k^{-(3/4)+\de}\quad\text{when}\quad k^{1/2}z>1.\cr}
\eqno (7.16)
$$
Indeed, if $k^{1/2}z<1$ then
$$\eqalign{
\Sl(l^2k)^{-(3/4)+\de}|\sin(\pi^{1/2}zlk^{1/2}/2)|
=\De^{(1/2)-2\de}k^{-(3/4)+\de}\Sl |l&\De|^{-(3/2)+2\de}|\sin(l\De)|\De,\cr
&\De=\pi^{1/2}zk^{1/2}/2,\cr}
$$
which proves the first part of (7.16). The second part follows from
the evident inequality
$$
\Sl(l^2k)^{-(3/4)+\de}|\sin(\pi^{1/2}zlk^{1/2}/2)|
\le k^{-(3/4)+\de}\Sl l^{-(3/2)+2\de}.
$$
Similarly,
$$
\sigma_{z\a}^2(k)=\int_0^1|\fk|^2ds\le
C_\de \Sl(l^2k)^{-(3/2)+2\de}\sin^2(\pi^{1/2} zlk^{1/2}/2),
$$
hence
$$
\sigma_{z\a}^2(k)\le\quad\eqalign{
&C_0(\de) k^{-1/2}z^{2-4\de} \quad\text{when}\quad k^{1/2}z<1,\cr
&C_0(\de) k^{-(3/2)+2\de}\quad\text{when}\quad k^{-1/2}z>1.\cr}
\eqno (7.17)
$$
Comparing (7.13) with (7.17), we obtain (7.14).

Let us prove (7.15). By (7.13) and (7.16), the
inequality
$$
\sup_{0\le s\le 1} |\fk|\ge (\Var \xi_{z\a})^{1/2}
$$
can hold only if
$$
C_0(\de)\, z^{(1/2)-2\de} k^{-1/2}\ge C\ep z^{1/2}|\log z|^{-1/2},
$$
which implies
$$
k\le z^{-6\de},\quad z\le z_0(\a,\de,\ep). \eqno (7.18)
$$
 From (7.15), (7.13) and (7.18),
$$
g_{z\a}(\ep)\le z^{-1-\de}\sum_{k=1}^{z^{-6\de}}\sigma_{z\a}(k),\quad z\le
z_1(\
   a,\de,\ep),
$$
so by (7.17),
$$
g_{z\a}(\ep)\le z^{-1-\de}\sum_{k=1}^{z^{-6\de}} k^{-1/2}z^{2-5\de}
\le z^{1-20\de},\quad z\le z_2(\a,\de,\ep),
$$
which proves that $\di\lim_{z\to 0}g_{z\a}(\ep)=0$.

Thus the Lindeberg condition (7.14), (7.15) holds,
and so the distribution of $(\Var \xi_{z\a})^{-1/2}\xi_{z\a}$
 converges to a standard normal distribution
as $z\to 0$. Theorem 1.11 is proved.

{\bf Appendix.} {\it Proof of the formula (4.2).}
 We prove in this Appendix the formula
$$
I=\pi^{-1}\II \II |\xi|^{-3}\rho (\xi )\sin^2 Y(\xi )d\xi =
\text { Area \{ Int }\g\} , \tag A.1
$$
which reduces to (4.2) when $\text { Area \{ Int }\g \} =1. $
 To prove (A.1) let us rewrite
the expression $I$ in polar coordinate system.
Given an angle $\f$ let $\xi (\f) $ be the unit vector
in $\R^2$ with this angle, and $\rho_0 (\f)=\rho (\xi(\f))$
and $Y_0(\f)=Y(\xi (\f))$.

Then
$$
\align
I&=\pi^{-1}\int_0^\infty\int_0^{2\pi}
r^{-2}\rho_0(\varphi)\sin^2(rY_0(\varphi))\,d\varphi
\,d\rho\\
&=\pi^{-1}\int_0^{2\pi}\rho_0(\varphi)
Y_0(\varphi)\,d\varphi\int_0^\infty\frac{\sin^2
u}{u^2}\,du=\frac12\int_0^{2\pi}\rho_0(\varphi)Y_0(\varphi)
\,d\varphi\;. \tag A.2
\endalign
$$
Given a non-zero vector $\xi$ in $\bold R^2$ with angle $\varphi$  let
$\psi(\varphi)$ be the angle
of the point $x=x(\f)\in\gamma$ for which
the outer normal is parallel to $\xi$.
Define the sector
$$
V(\psi_1,\psi_2)=\{x\in\bold R^2\: \quad\psi_1<\text{the angle of the
vector}\; x<\psi_2\}\;.
$$
Then
$$
\rho_0(\varphi_1)Y_0(\varphi_1)(\f_2-\f_1)=2\text{Area}\{ \,\left(V(
\psi(\varphi_1);\psi(\varphi_2))\cap\text{Int}\,\gamma\right) \}
+o(|\varphi_1-\varphi_ 2|)\;.
$$
The last relation holds, since the set
$V(\psi(\varphi_1);\psi(\varphi_2))\cap \text{Int}\, \gamma$ can be well
approximated by a triangle with the baseline
$\De s =(\f_2-\f_1)\rho_0(\f_1)$
and the height  $Y_0(\varphi)$.
Hence, we get, by approximating the integral by the usual
approximating sum that
$$
\int_0^{2\pi}\rho_0(\varphi)Y_0(\varphi)\,d\varphi=2\text{Area}\,\text{Int}\,\ga
   mma
\;.
$$
The last identity together with (A.2) imply formula (A.1).

{\it Acknowledgements.} The authors are grateful to Freeman Dyson
and Peter Major for useful remarks.
They are also indebted to Peter Major for a short proof of formula (4.2).
The first author thanks the Institute for Advanced Study, Princeton,
and Rutgers University for financial support. The work was also supported by
the
    Ambrose Monell Foundation
and by NSF Grant 89--18903.

\beginsection References \par

\item {[B1]} {M. V. Berry, Semiclassical mechanics of regular and irregular
moti
   on,
In {\it Chaotic behavior of deterministic systems}, Les Houches Lectures, vol.
X
   XXVI,
ed. G. Iooss, R. H. G. Helleman \& R. Stora, 171--271. Amsterdam:
North-Holland,
1983.}

\item {[B2]} {M. V. Berry, Semiclassical theory of spectral rigidity,
{\it Proc. Roy. Soc. London}, Ser. A, {\bf 400}, 229--251 (1985).}

\item {[BT]} {M. V. Berry and M. Tabor, {\it Level clustering in the regular
spe
   ctrum},
{\it Proc. Roy. Soc. London}, Ser. A, {\bf 356}, 375--394 (1977).}

\item {[Bes]} A. S. Besicovitch, {\it Almost periodic functions}, Dover
Publications, New York, 1958

\item {[Bl1]} {P. M. Bleher, Quasiclassical expansion and the
problem of quantum chaos, {\it Lect. Notes in Math.} {\bf 1469},
60--89 (1991).}

\item {[Bl2]} {P. M. Bleher, On the distribution of the number of
lattice points inside a family of convex ovals, {\it Duke Math. Journ.}
{\bf 67}, 3, 461--481 (1992).}

\item {[Bl3]} {P. M. Bleher,
Distribution of the error term in the Weyl asymptotics for the Laplace
operator on a two--dimensional torus and related lattice problems,
Preprint, Institute for Advanced Study, IASSNS-HEP-92/80, 1992
(to appear in {\it Duke Math. Journ.}).}

\item {[BCDL]} P. M. Bleher, Zh. Cheng, F. J. Dyson and J. L. Lebowitz,
Distribution of the error term for the number of lattice points
inside a shifted circle, Preprint, Inst. Adv. Study,
IASSNS-HEP-92/10, 1992 (to appear in {\it Commun. Math. Phys.}).

\item {[BD1]} P. M. Bleher and F. J. Dyson,
Mean square value of exponential sums related to representation of integers as
s
   um of
two squares,
Preprint, Inst. Adv. Study, IASSNS-HEP-92/84, 1992.

\item {[BD2]} P. M. Bleher and F. J. Dyson, The variance of the error
function in the shifted circle problem is a wild function of the shift,
Preprint, Inst. Adv. Study, IASSNS-HEP-92/83, 1992.

\item {[BGGS]} E. B. Bogomolny, B. Georgeot,
M.-J. Giannoni and C. Schmit,
Chaotic billiards generated by arithmetic groups,
{\it Phys. Rev. Lett.} {\bf 69}, 1477--1480 (1992).

\item {[Boh]} O. Bohigas, Random Matrix Theory and Chaotic
Dynamics, {\it Les Houches, Session LII, 1989}, Eds.
M.-J. Giannoni, A. Voros and J. Zinn-Justin, Elsevier
Sci. Publ., 89--199, 1991.

\item {[CCG]} {G. Casati, B. V. Chirikov, and I. Guarneri,
Energy--level statistics of integrable quantum systems,
{\it Phys. Rev. Lett.} {\bf 54}, 1350--1353 (1985).}

\item {[CGV]} G. Casati, I. Guarneri, and F. Valz-Gris,
Degree of randomness of the sequence of eigenvalues,
{\it Phys. Rev.} {\bf A 30}, 1586--1588 (1984).

\item {[CdV]} {Y. Colin de Verdi\`ere, Nombre de points entiers
dans une famille homoth\`etique de domaines de $\R^n$,
{\it Ann. Scient. \'Ec. Norm. Sup.}, $4^e$ s\'erie,
{\bf 10}, 559--576 (1977).}

\item {[CFS]} {I. P. Cornfeld, S. V. Fomin
and Ya. G. Sinai, {\it Ergodic Theory}, Springer, New York, 1982.}

\item {[CL]} {Z. Cheng and J. L. Lebowitz, Statistics of energy
levels in integrable quantum systems, {\it Phys. Rev.}
{\bf A44}, R3399--R3402 (1991).}

\item {[CLM]} Z. Cheng, J. L. Lebowitz, and P. Major,
On the number of lattice points between two enlarged
and randomly shifted copies of an oval,
(in preparation).

\item {[GH]} H.-D Gr\"af, H. L. Harney, H. Lengeler,
C. H. Lewenkopf, C. Rangacharyulu, A. Richter,
P. Schardt, and H. D. Weidenm\"uller,
Distribution of Eigenmodes in a Superconducting Stadium Billiard
with Chaotic Dynamics, {\it Phys. Rev. Lett.} {\bf 69},
1296--1299 (1992)

\item {[Gu]} M. C. Gutzwiller, {\it Chaos in classical and quantum
mechanics}, Springer, N. Y. e.a., 1990.

\item {[HW]} {G. H. Hardy and E. M. Wright, {\it An introduction to the
theory of numbers}, 4th Edn, Oxford, 1960.}

\item {[H-B]} {D. R. Heath-Brown, The distribution and moments of the
error term in the Dirichlet divisor problem, Preprint, Oxford
University (to appear in {\it Acta Arithmetica}).}

\item {[K]} {D. G. Kendall, On the number of lattice points inside a
random oval, {\it Quart. J. of Math.} (Oxford) {\bf 19}, 1--26 (1948).}

\item {[L]} M. Lo\`eve, {\it Probability theory}, 4th Edn., Springer,
N. Y. e.a., 1977.

\item {[M]} P. Major, Poisson law for the number of lattice
points in a random strip with finite area,
{\it Probab. Theory Relat. Fields} {\bf 92}, 423--464 (1992).

\item {[Me]} M. L. Mehta, {\it Random matrices}, 2nd Edn., Academic Press,
Boston e.a., 1991.

\item {[R]} B. Randol, A lattice--point problem, {\it Trans. Amer. Math. Soc.}
{\bf 121}, 257--268 (1966).

\item {[LS]} W. Luo and P. Sarnak, Number variance for arithmetic
hyperbolic surfaces (in preparation).

\item {[Sar]} P. Sarnak, Arithmetic quantum chaos, Schur Lectures,
Tel Aviv, 1992.

\item {[S1]} Ya. G. Sinai, Poisson distribution in a geometric
problem, In: {\it Advances in Soviet Mathematics}, v. 3,
ed. Ya. G. Sinai, AMS, Providence, 199--214, 1991.

\item {[S2]} {Ya. G. Sinai, Mathematical problems in the theory of
quantum chaos, {\it Lect. Notes in Math.} {\bf 1469}, 41--59 (1991).}

\vfill\eject

\beginsection  Figures Captions \par

Fig. 1. Error function of the circle problem.

Fig. 2. Scaling function of the circle problem.

Fig. 3. Scaling function for an ellipse with $a_1/a_2=1/\pi$,
$\a=(0.,0.)$.

Fig. 3. Scaling function for an ellipse with $a_1/a_2=1/\pi$,
$\a=(0.1,0.1)$.

\bye